# Aqueous self-assembly of a wide range of sophorolipid and glucolipid microbial bioamphiphiles (biosurfactants): considerations on the structure-properties relationship

Niki Baccile[a,*] Alexandre Poirier,[a] Patrick Le Griel,[a] Petra Pernot,[b] Melike Pala,[c] Sophie Roelants,[d,e] Wim Soetaert[d,e] Christian V. Stevens[c]

[a] Sorbonne Université, Centre National de la Recherche Scientifique, Laboratoire de Chimie de la Matière Condensée de Paris, LCMCP, 4, Place Jussieu, F-75005 Paris, France
[b] ESRF – The European Synchrotron, CS40220, 38043 Grenoble, France
[c] Ghent University, Faculty of Bioscience Engineering, Department of Green Chemistry and Technology, Coupure Links 653, B-9000 Gent, Belgium
[d] Centre for Industrial Biotechnology and Biocatalysis (InBio.be), Department of Biotechnology, Faculty of Bioscience Engineering, Ghent University, Coupure Links 653, 9000 Ghent, Belgium
[e] Bio Base Europe Pilot Plant, Rodenhuizenkaai 1, 9042 Ghent, Belgium

* Corresponding author:
Dr. Niki Baccile
E-mail address: niki.baccile@sorbonne-universite.fr
Phone: +33 1 44 27 56 77

**Keywords**

Bioamphiphiles; Biosurfactants; Phase behaviour; Self-assembly; SAXS; Structure-properties relationship

**Abstract**

Sophorolipids are well-known scaled-up microbial glycolipid biosurfactants with a strong potential for commercialization due to their biological origin and mildness in contact with the skin and the environment compared to classical surfactants. However, their association properties in water are still poorly understood, they cannot be predicted and their behavior in solution challenges half a century of knowledge generated in the field of surfactant science. By studying forty different types of sophorolipids and sophorosides in water using small angle X-ray scattering, and optical and cryogenic transmission electron microscopy, this work provides better understanding of their structure-property relationship and identifies which chemical groups in their molecular structure have a critical influence on their self-assembly properties. Structural features like the number of sugar headgroups, acetylation, end-chain functional group, (un)saturation, lactonization and length of chain are adjusted to both rationalize their impact and understand their effect on self-assembly. The number of sugar groups, pH, (un)saturation and lactonization were found to have a critical impact on sophorolipid self-assembly. The chemical nature of the end-chain functional group and chain length were also found to have a possibly critical impact, depending on the specific type of chemical function (COOH and long chains are critical). Mono- and diacetylation, as well as the position of sophorose in the fatty acid (ω, ω-1), are not critical, i.e., they did not significantly influence sophorolipid self-assembly.

**Introduction**

Biosurfactants are biological amphiphiles derived from natural resources and obtained by plant extraction, enzymatic synthesis, or microbial fermentation. The latter certainly gives rise to one of the largest families and has been known for decades[1–4]. It contains both glycosylated (sophorolipids, rhamnolipids, mannosylerythritole lipids, etc.) and peptidic (surfactin) lipids. Due to their lower environmental impact, microbial amphiphiles have been developed for decades to replace synthetic surfactants.[2–4] Microbial biosurfactants are indeed considered to be more biodegradable and less toxic than petrochemical surfactants, and therefore they are used in a number of applications in detergency, cosmetics, environmental science, or as antimicrobial compounds,[5–8] with a milder effect on protein denaturation.[9]

Although great effort has been dedicated in recent years to bring microbial biosurfactants to the market,[10–14] also stimulated by increasing interest from industry,[15,16] their effective use in real-life applications will certainly depend on the knowledge of their physicochemical properties in water under diluted and semi-diluted conditions (typically between 0.1 and 20 wt%), classically representing surfactant concentrations in commercial products. Although surface tension, critical micelle concentration (cmc) and solubility are classically studied when biosurfactants are produced,[17] more advanced studies on their phase behavior are also needed.

The study of biosurfactant self-assembly beyond the cmc is a relatively new topic of research[18] and reviewed by us recently.[19] According to the present state of the art, most studies exploring concentrations below 20 wt% concern surfactin, acidic and lactonic C18:1 sophorolipids, mannosylerythritol lipids (MELs) and rhamnolipids. At higher concentration, the number of works is even lower and mainly concerns MELs[20–25] and sophorolipids.[26] Although these molecules certainly represent the most important biosurfactants developed for decades,[12,17] little is known about the new biosurfactants developed during the past few years from both chemo-enzymatic modifications[27] and genetic engineering.[10,28,29]

Sophorolipids, SLs, are certainly the largest group of biosurfactants in terms of derivatization. This is certainly explained by the importance of sophorolipids in the biosurfactant panorama, as SLs can be prepared in large amounts and thus be easily upscaled. They have antimicrobial, antiviral and anticancer properties, and can be used to stabilize nanomaterials and cosmetic formulations.[10,30–33] The second important feature of SLs is the presence of two reactive functions, namely the *cis* double bond and carbonyl (acid or ester) group. These have been the obvious targets of most molecular derivatization approaches.[27] The third and probably most important feature is the decoded genome of *S. bombicola*, the microorganism producing sophorolipids, thus paving the way for controlled metabolic engineering aimed at transforming this exotic microorganism into an important platform yeast.[34,35] Although recent work has focused on the self-assembly properties of sophorolipid derivatives obtained by chemical[36] or genetic (e.g., glucose lipids[37] or acetylated sophorolipids[38]) engineering, nothing is known about the much broader family of sophorolipid derivatives.

The data existing so far show that the solution behaviour beyond the critical micelle concentration of sophorolipids in particular, and biosurfactant amphiphiles in general, cannot be rationalized according to the general predictive theories of amphiphile self-assembly, like that based on the packing parameter, known since the 1970s,[39] and those of its variants that have been implemented.[40–42] In a recent review,[19] we have shown that the risk of failure is high, when attempting to adapt the packing parameter approach to predict and understand the self-assembly of sophorolipids and biosurfactants.

To move forward, a more empirical approach is needed to help develop a new theoretical framework, better adapted to this class of compounds. This work therefore explores the self-assembly properties at room temperature of diluted and semi-diluted (< 20 wt%) solutions of 40 different sophorolipids, obtained either by fermentation or by chemical derivatization. By doing so, it is possible to elucidate the effect of acetylation, (un)saturation, type of end-group, chain length, $\omega$ and the $\omega$-1 position of the acetal bond. Each structural element will be categorized in terms of its impact (critical, possibly critical, non-critical) on the self-assembly of sophorolipids. Considering the fact that a general theory of amphiphile self-assembly able to predict, or even explain, the behavior of complex molecules like sophorolipids does not yet exist, this work therefore constitutes the first attempt to establish an empirical structure-property relationship, to fill a gap in knowledge regarding sophorolipids and other complex, functional, amphiphiles.[43]

**Material and methods**

*Chemicals.* All the glycolipids were produced in the framework of the Applisurf project (http://www.bbeu.org/pilotplant/applisurf) and were provided by the Bio Base Europe Pilot Plant, Gent, Belgium, the Department of Green Chemistry and Technology, Ghent University, and Amphistar, Belgium. As a general strategy, developed in previous publications,[44,45] molecular identification is performed by high-performance liquid chromatography (HPLC) equipped with an evaporative light scattering detector (ELSD), liquid chromatography (LC) coupled to mass spectroscopy (MS), solution nuclear magnetic resonance (NMR) and, sometimes, ultraviolet (UV) and infrared (IR) spectroscopies. Glucose, glycerine and organic acids are typically determined by HPLC equipped with a refractive index detector (RID); free fatty acids or oil are detected by thin layer chromatography (TLC) or gas chromatography equipped with (MS); organic nitrogen or proteins by bicinchoninic acid assay (BCA) and/or the external Kjeldahl method; and, finally, ash content is determined after thermal treatment. In particular, purity is systematically verified using HPLC and/or nuclear magnetic resonance (NMR). Overall, the compounds have a reduced content of free fatty acids and congeners, with a certified purity on average above 95%. Samples are numbered from (**1**) to (**40**) and they are presented in groups according to structural affinities in specific sections in the main text. All the molecular structures are also presented in Supporting Information (see *Sample analysis* paragraph below) and summarized in Table S 1.

More specifically, chemical and structural analyses for sample (**1**) are published in Ref. 44, for samples (**3**), (**4**) and (**5**) in Ref. 45, for sample (**12**) in Ref. 46, for samples (**7**), (**8**) and (**9**) in Ref. 47 and samples (**14**), (**15**) and (**17**) in Ref. 48. The synthesis procedure and structural analysis of samples (**6**), (**29**), (**30**), (**31**), (**32**), (**33**), (**34**), (**35**), (**36**), (**37**), (**38**), (**39**), (**40**) are given in Supporting Information (Page S2-S10). The synthesis procedure and structural analysis of samples (**2**), (**10**), (**11**), (**13**), (**16**), (**18**), (**19**), (**20**), (**21**), (**23**), (**24**), (**25**), (**26**), (**27**), (**28**) are reported elsewhere.[49]

*Solution nuclear magnetic resonance (NMR).* $^1$H NMR and $^{13}$C NMR spectra were recorded at 25 °C at 400 and 100 MHz, respectively, using a Bruker Avance III-400 spectrophotometer. The samples were dissolved in deuterated solvents as indicated with the spectral data of each compound. All spectra were processed using TOPSPIN 3.2 and acquired through the standard sequences available in the Bruker pulse program library. The peaks were assigned by using classical 2D spectra (COSY, HSQC, HMBC and/or H2BC), previously employed for sophorolipid and sophoroside identification.[44,45] TMS is used as usual reference for chemical shift (0 ppm).

*Sample preparation.* The samples are prepared by dilution from concentrated solutions. Samples at 5, 2.5 and 1.25 mg/mL are prepared from a sample solution having a concentration of 10 mg/mL. Solutions at 100, 50 and 25 mg/mL are prepared from a solution having a sample concentration of 200 mg/mL. MilliQ-grade water is used as such for all solutions. The corresponding weight fractions are, from the lowest to the highest, 0.12, 0.25, 0.5, 1.0, 2.4, 4.8, 9.1 and 16.7 wt%. Solutions are vigorously vortexed at room temperature until solubilization, although in some cases, heating at 80°C using a dry bath and vortexing is necessary to either improve solubilization or colloidal dispersion.

*Solubility* Surfactant solubility was tested at room temperature in mQ water buffered to pH 4, 6 and 8 ($KH_2PO_4$/NaOH/$H_2O$ buffer). Preliminary solubility tests were done at 0.1 wt% according to the following protocol. Surfactant powder was weighed in 40 mL glass vials using a SWING automated reaction and sample preparation deck (Chemspeed Technologies AG). After solvent addition, vials were placed on a roller (ROLLER 10 digital, IKA) for 24h at 50 rpm to ensure sufficient mixing. All the samples were then split and assigned to the following classes based on visual inspection: soluble (clear solution), turbid (not clear) and insoluble (sedimented) samples. Turbid samples were stored without shaking at room temperature for 24 h and categorized as insoluble if sedimentation eventually occurred. If no sedimentation was observed, the surfactant was categorized as soluble. The same test was performed at 2 wt% for all the surfactants categorized as soluble. If the surfactants remained soluble at the latter concentration, additional surface tension measurements were performed. Additional solubility tests are performed for surfactants that are soluble at 0.1 wt% and insoluble at 2 wt% by incrementing the concentration by 0.1 wt% until maximal solubility was reached.

*Surface tension and critical micelle concentration (CMC)* Surface tension experiments were performed by Flamac, Belgium, using an automated approach. Pendant drop shape analysis, using a Surface Energy Screening (SES) platform (First Ten Angstroms FTA-2000) and accompanying FTA software, was used to optically measure the surface tension of the surfactant solutions in mQ water at 2; 1; 0.5; 0.2; 0.1; 0.05; 0.02; 0.01; 0.005; 0.001; 0.0001 wt%. The samples were loaded onto the SES platform with increasing concentration and the automated measurement of the dilutions was done as described next. 20 µL of surfactant solution was aspirated from the first vial with the lowest concentration. During the retraction of the 25G Teflon® tip from the vial a small amount of air was aspirated (2 µl) to prevent dripping. The tip was then brought into the measurement position, a drop (6 µL) formed and optical image was taken before the next sample was measured. The resulting optical image was analyzed and the resulting surface tension values used to determine the critical micelle concentration (CMC). A plot of surface tension as a function of the logarithm of surfactant concentration was generated. The CMC was determined by the intersection between the linear fit through the concentration-dependent region and the linear fit passing through the concentration-independent plateau at higher surfactant concentrations. Typical surface tension plots with related linear fits are shown in Figure S 1 in Supporting Information, while Table S 2 provides the raw surface tension data of the entire set of compounds.

*Small angle X-ray scattering (SAXS).* SAXS experiments for the samples below 16 wt% were performed using the automatic sample changer available at the BM29 BioSAXS beamline (Proposal N° MX2311).[50,51] The energy was E = 12.5 KeV and a sample-to-detector distance of 2.83 m. q was the wave vector, with q = 4π/λ sin(θ), 2θ corresponding to the scattering angle and λ the wavelength. The q-range was calibrated between ∼ 0.05 < q / nm$^{-1}$ < ∼ 5, readily available raw data obtained on the 2D detector were integrated azimuthally using the in-house software provided at the beamline and thus obtain the typical scattered intensity I(q) profile.[52] Absolute intensity units were determined by measuring the scattering signal of water ($I_{(q=0)}$ = 0.0163 cm$^{-1}$) at high q-values, where it was independent of the wavevector, q. All the SAXS profiles are shown in Supporting Information while the entire set of the corresponding *.dat files are grouped, zipped and freely available for further use at https://doi.org/10.5281/zenodo.7677356

Given the considerable amount of data, the differences between and the coexistence of the type of phases, the analysis of the SAXS data is performed with model-independent Guinier (Eq. 1)

$$I(q) = I(0) e^{\left[\frac{-q^2 R_g^2}{3}\right]}, q R_g < 1 \qquad \text{Eq. 1}$$

and power law (Eq. 2)

$$I(q) \propto q^{-m} \qquad \text{Eq. 2}$$

analyses for, respectively, the mid-*q* and low-*q* range, with $R_g$ = radius of gyration and *m* > 0 being a dimensionless exponent. The Guinier analysis is performed by fitting the plateau of the SAXS data at q< 1 nm$^{-1}$, generally so that $qR_g < 1$, directly using Eq. 1 using SasView 3.1.2 software (Guinier model, scale and Rg being the only two free parameters). The relationship between Rg and the sphere of a radius, R, is given by Eq. 3

$$R = \sqrt{\frac{5}{3} Rg^2}. \qquad \text{Eq. 3}$$

*Microscopy*. Microscopy experiments were performed using several instruments according to the resolution required. Ultra-high resolution was achieved using a cryogenic transmission electron microscope (cryo-TEM, FEI Tecnai 120 twin microscope operated at 120 kV and equipped with a Gatan Orius CCD numeric camera). Cryofixation was performed on a home-made cryo-fixation device. The solutions were deposited on glow-discharged holey carbon coated TEM copper grids (Quantifoil R2/2, Germany). Excess solution was removed and the grid was immediately plunged into liquid ethane at -180°C. All the grids were kept at the temperature of liquid nitrogen throughout all the experiments.

Lower resolution was obtained with an optical microscope equipped with polarized light (transmission Zeiss AxioImager A2 POL optical microscope equipped with an AxioCam CCD camera).

**Results**

All the sample solutions studied in this work were analyzed by small-angle X-ray scattering, and in the case of ambiguity in attributing the nature of the self-assembled structure, supplementary optical microscopy, polarized light microscopy or cryo-TEM experiments were performed on the samples concerned. Given the considerable amount of data, and the differences between and the coexistence of the types of phases, the analysis of the SAXS data was performed with a model-independent approach. This was also justified by the fact that the coexistence of more than one phase was often observed, thus making the systematic use of a model-dependent approach too complex and unreliable.

The value of the slope ($m$ in Eq. 2) at low q (generally at q< 0.1 nm$^{-1}$) in the Log(I)-Log(q) representation is widely known to be related to the morphology of the scattering objects.[53] On the one hand, those SAXS profiles characterized by a flat plateau at low-q and an oscillation at high-q can safely be attributed to spheroidal micelles.[54] On the other hand, strong low-q scattering characterized by a slope having typical values of -1 (rods)[53] and -1.6 (rods in a good solvent)[55,56] are generally attributed to cylindrical and wormlike micelles, respectively. The slope values of -2 (flat membranes)[53] could, on the contrary, be attributed to either bilayers (vesicles or lamellae) or fibers with a flat cross section. The difference between a bilayer or a fiber is generally determined by the presence of a broad diffraction peak often above 1 nm$^{-1}$ (see for instance SAXS of fibrillar sophorolipids in Ref. [57,58]). The differences between bilayers, fiber and micellar morphologies are generally obtained by means of cryo-TEM, as presented for instance in Ref. [59]. The typical reference SAXS profiles, with related slope ($m$ in Eq. 2) values at low-q, simulated from sphere, cylinder or membrane form factor models and repulsive hard sphere interactions on sphere model are given in Figure 1a,b. The typical experimental SAXS profiles for crystalline self-assembled fibers (obtained from glucolipids[58]) are shown in Figure 1c.

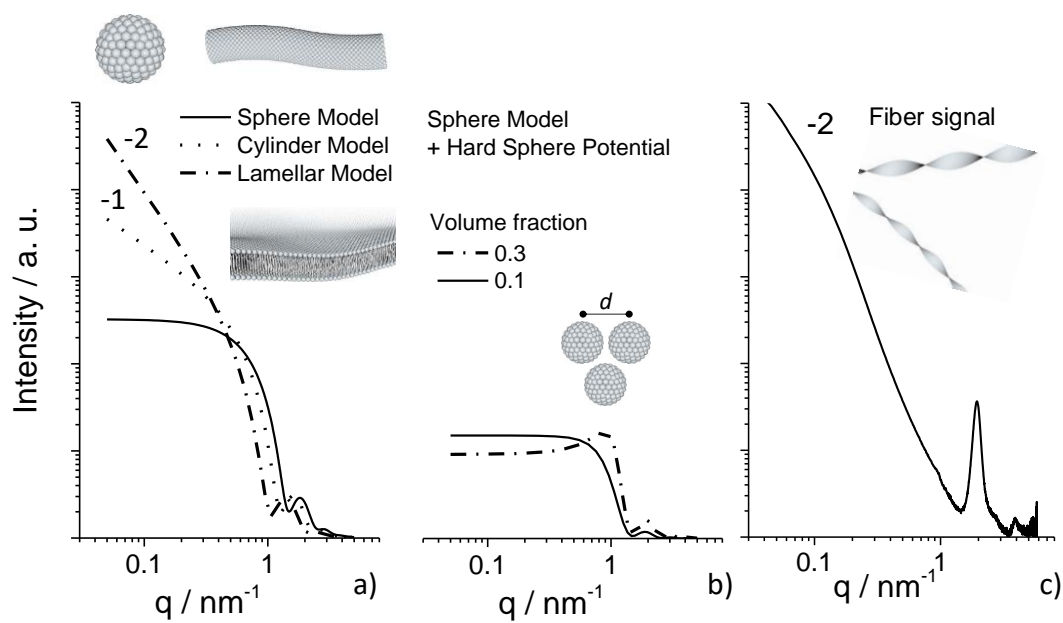

**Figure 1 – Typical SAXS profiles associated with spherical, cylindrical, lamellar and fibrous morphologies as well as typical profiles found for high volume fractions of spherical objects. Simulated SAXS profiles generated with SasView 3.1.2 software are shown in a) for sphere, cylinder, lamellar form factor models and in b) for a combination of sphere form factors with hard sphere structure factors. Typical realistic parameters for all simulations are: background (0.001 cm$^{-1}$), scale (1), radius (3 nm), radius polydispersity (0.1), scattering length density of solvent (water, 9.4·10$^{-4}$ nm$^{-2}$) and object (11·10$^{-4}$ nm$^{-2}$), length of cylinder (500 nm), thickness of membrane (6 nm), volume fraction of hard sphere potential (0.1, 0.3). c) Typical experimental scattering profile associated with self-assembled crystalline fibers, previously reported for analogous glycolipid samples.[58]**

Whenever possible, the radius of gyration, $R_g$, was estimated from SAXS data using the Guinier analysis according to Eq. 1.[53] Although more qualitative than a full-model analysis, the Guinier approach is still reliable and has the advantage of comparing large amounts of data without too many hypotheses on morphology. Finally, some systems can undergo repulsive interactions, originating from high volume fractions or the presence of charges.[54] In this case, a structure factor changes the SAXS profiles of Figure 1a. If the volume fraction is small (e.g., around 0.1), the SAXS profiles are flattened out compared to the structure factor-free profile, as shown in the example of a sphere in Figure 1b. If the volume fraction is higher, a broad structure peak (**BSP**) is generally observed[54], as illustrated in the example in Figure 1b for a volume fraction of 0.3. Occasionally, sharp peaks, often in a 1:2:3… ratio, indicate the crystallinity of the sample and their position is systematically indicated. These are referred to as sharp lamellar peaks, **SLP**, when they superimpose a SAXS profile typical of a lamellar morphology. Otherwise, they are simply referred to as **CP**, crystalline peaks, when they overlap with a non-lamellar signal. If the organization of the lamellae is loose, the peaks are rather broad and referred to as broad lamellar peaks, **BLP**, similar to what are generally found for swollen lamellar phases.[60,61] Finally, the critical micellar concentrations (CMCs) are provided for most compounds in Table S 1. The comparison between CMCs and the lowest concentration studied in this work (0.12 wt%) shows that, except for several C9 derivatives of sophorolipids displaying high CMC values, all the compounds are studied above their respective CMC. As a general trend, C9 derivatives are more soluble than most other compounds and this fact will be discussed specifically in the section dedicated to this family of molecules.

In the following section, the self-assembly under semi-diluted conditions (< 16.7 wt%) of a broad set of biosurfactants will be discussed by family. In terms of nomenclature, the term sophorolipid (SL) is always used for molecules containing an acid (COOH) or an ester (COOR) group, while sophoroside (SS) is generally used for molecules containing an acetal bond between the glucose and the alkyl chain and a non-carboxylic (or ester) end group. Compounds are also rationalized in terms of the position of the acetal bond on the aliphatic chain, [ω] (last) or [ω-1] (second to last), number of unsaturation, C18:1 (monounsaturated) or C18:0 (saturated), and chain length, C18 or C9. A general discussion about the effects of acetylation (acetyl vs. non-acetyl, Table S1), structure (open acidic or lactonic), number of glucose groups, chain length and ionization will also be proposed.

**Self-assembly of bioamphiphiles by family**

*Bolaform sophorolipids/sophorosides*, (**1**)-(**5**) in Figure 2. This class of biosurfactants is characterized by a pseudo-symmetry in the distribution of the sophorose headgroups and iis not sensitive to pH over a broad pH range, with the exception of compound (**1**), which can be hydrolyzed under strong alkaline conditions.

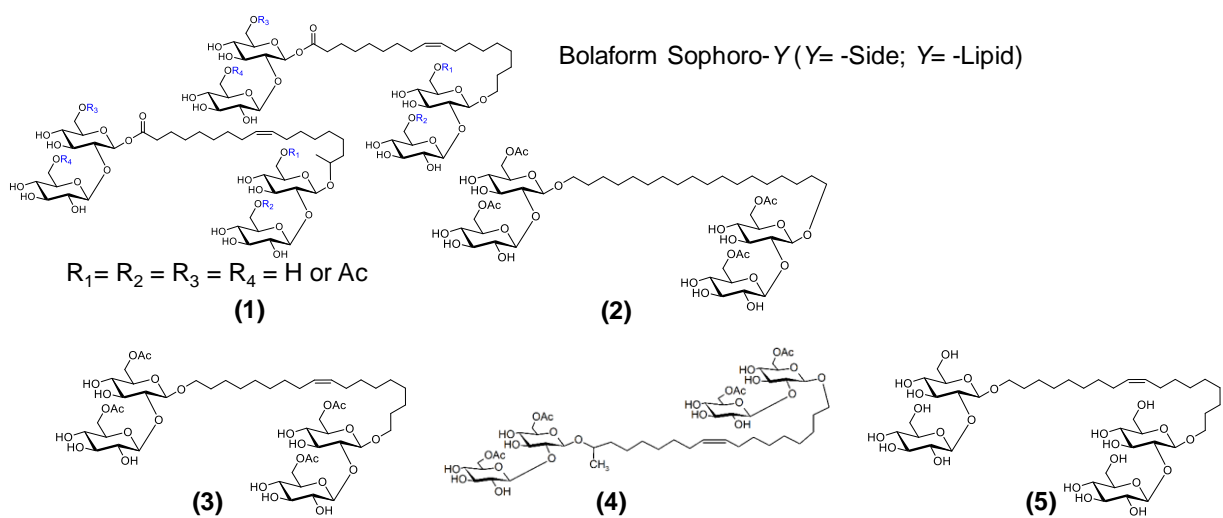

**Figure 2 – Chemical structures of the bolaform sophorolipids (SL) and sophorosides (SS) studied in this work.**

Although all bolaform SL and SS can be dissolved in water with low CMC, gentle warming (above 40°C) may be necessary for some of them (**1**), (**2**), (**5**). When a clear solution is obtained, bolaform SL and SS form a micellar phase (**M**). The typical SAXS profiles, and related typical fit of the Guinier region (Eq. 1 using SasView 3.1.2), are shown for (**1**), (**3**), (**4**) in Figure 3a at the selected concentration of 9.1 wt%, while the entire set of SAXS profiles of (**1**) to (**5**) at all the concentrations explored are given in Figure S 2. According to the Guinier analysis in the limits $qR < 1$, $R_g$ is in the order of 1.8 nm (Table 2), corresponding to a sphere of radius, R, of 2.3 nm (Eq. 3).

**Table 1 - Phase behavior of sophorolipids and sophorosides. The $N°$ column refers to compound numbers in Figure 2. M: Micelles, F: Fibers.**

| N° | Phase (Concentrations in wt%) | | | | | | | | Comments |
|---|---|---|---|---|---|---|---|---|---|
| | 0.12 | 0.25 | 0.5 | 1.0 | 2.4 | 4.8 | 9.1 | 16.7 | |
| (1) | | | | M * | | | | | Micelles after heating. Fiber precipitate in time |
| (2) | | | | M / F | | | | | (**M** + **F**) --> **F** with time. Gel or spherulites depending on cooling rate. |
| (3) | | | | M * | | | | | *: Fluffy fibrous aggregates can precipitate after days/weeks. |
| (4) | | | | M * | | | | | |
| (5) | | | | M / F | | | | | Micelles after heating. Monophase fibrillation and gel in time. |

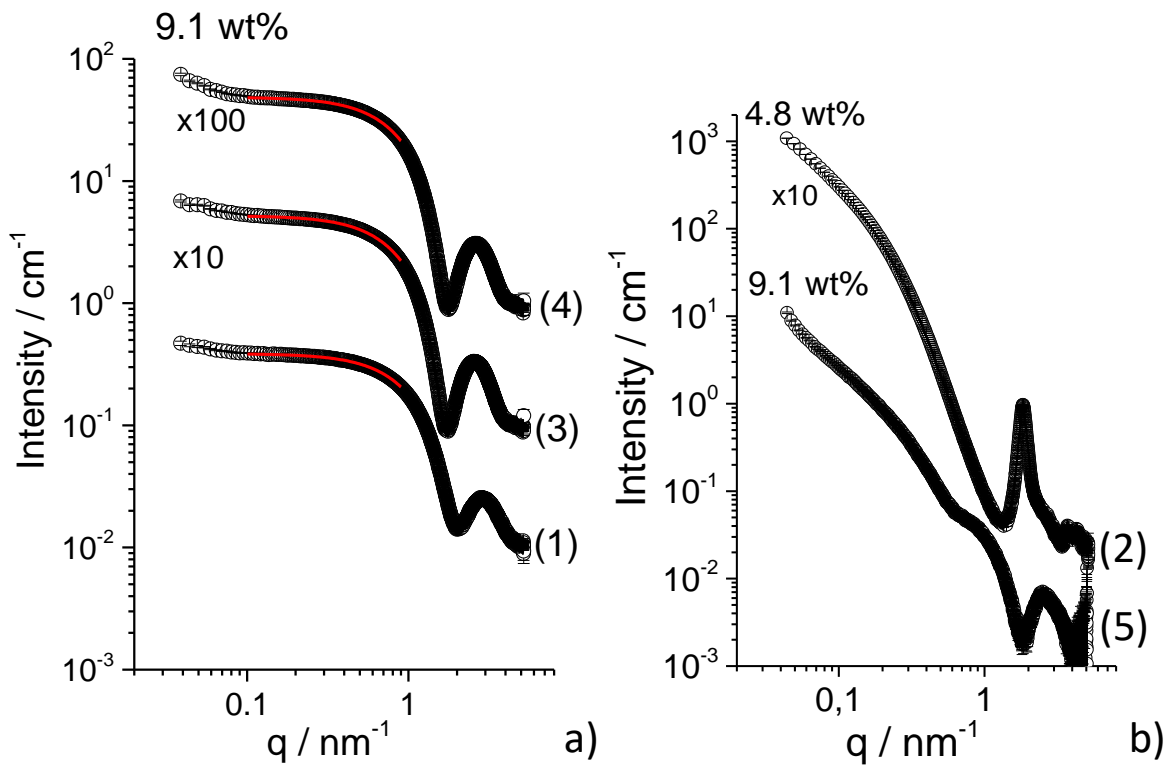

**Figure 3** – a) SAXS profiles of (1), (3) and ((4) at C= 9.1 wt% and related Guinier fit (SasView 3.1.2). The multiplication factors are given below each profile. b) SAXS profiles of (2) and (5). The concentrations and multiplication factors are given in the image.

Considering that the length of a disaccharide is about 1 nm[62] and that of a C18 chain is about 2.4 nm according to the Tanford formula[63] (Eq. 4)

$L$= 1.54+1.265*$n$                                                                                                              Eq. 4

with $L$ being the length of the alkyl chain and $n$ the number of $CH_2$ groups, the full length of a typical bolaform SL and SS is in the region of 4.4 nm. This value is approximately twice that of the micellar radius, R, thus suggesting that the micellar diameter contains a single bolaform SL or SS molecule, in agreement with theory on bolaform amphiphiles.[64]

All bolaform SL and SS samples may also be characterized by a second, fibrillar crystalline (**F**), phase, which forms over time. Of all the samples studied, the fiber phase is favored and thermodynamically stable at room temperature for (**2**) and (**5**). The fiber phase is corroborated by a series of evidence. Firstly, (**2**) and (**5**) form hydrogels at concentrations above 1 wt% and their elastic properties are reported elsewhere.[65,66] Their typical SAXS profile, given in Figure 3b (full set of profiles is given in Figure S 2), matches previous data[65,66] and cryo-TEM (Figure 4) confirms the fibrous nature of the samples.

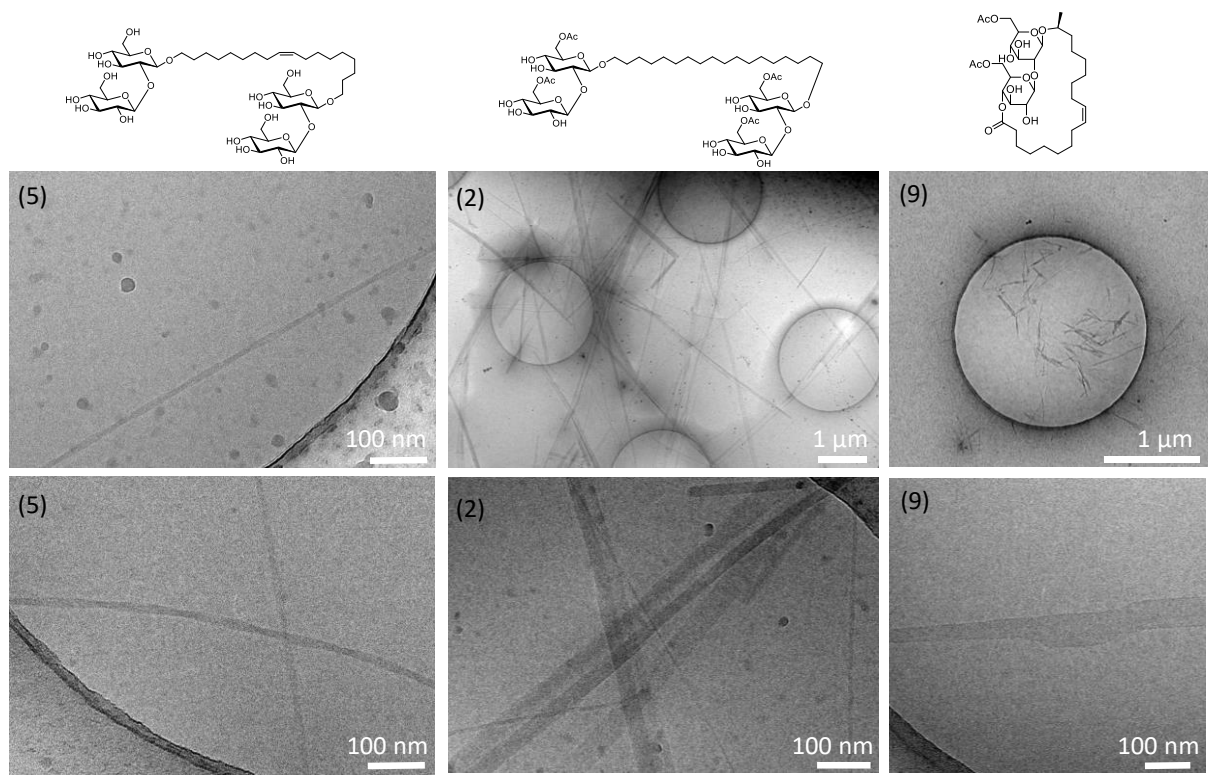

**Figure 4** - Cryo-TEM (C(**9**)= 0.25 wt%, C(**2**)= C(**5**)= 0.05 wt%) of aqueous solutions of sophorolipids.

On this basis, it must be added that fluffy fibrous aggregates may sediment after time for (**1**), (**3**) and (**4**). Although the micelle-to-fiber transition seems to be time-dependent at room temperature for all the samples, the kinetics is long, varying from days to weeks, and variable across samples. This parameter was not studied in this work, although it has been discussed for (**2**) elsewhere.[66]

**Table 2** – Parameters extracted from the SAXS data for sophorolipids and sophorosides (**1**) to (**5**) in Figure 2. CS: Clear solution, ColS: Colloidal solution, Rg : Radius of gyration, *: Fast heating-cooling (80°C → 25°C).

| N° | State | pH | Rg (nm) | Slope < 0.08 nm$^{-1}$ | Peaks (nm$^{-1}$) |
|---|---|---|---|---|---|
| (1) | CS | 4.4 | 1.6 | 0.2 | - |
| (2) | ColS/Gel* | 6.3 | - | 2.6 | - |
| (3) | CS | 6.0 | 1.9 | 0.2 | - |
| (4) | CS | 6.6 | 1.8 | 0.4 | - |
| (5) | Gel | 6.8 | - | 1.6±0.1 | 2.39 |

Finally, it is remarkable that micelles keep their spheroidal morphology across all concentrations and up to 16 wt% (Table 1, Table 2). This result is in contrast with the classical behavior of head-tail surfactants in water and for which concentration generally drives a sphere to rod-, or worm-like morphological transition, generally associated with a second CMC.[67] This means that adding more and more bolaform amphiphile to water does not change the shape (spheroidal) but it increases the volume fraction of the micelles in solution. This also occurs for (**1**), (**3**), (**4**) and (**5**) even at concentrations higher than 16 wt%. In a previous study[26] carried out at concentrations up 90 wt% using a combination of polarized optical microscopy and SAXS, these same compounds presented an isotropic phase up to about 60-70 wt%, with a clear-cut columnar hexagonal phase identified for (**3**) and (**4**) above 60-70 wt%.

The driving force for the **M**-to-**F** phase transition remains poorly understood, although it could be related to the minima of the conformation energy of the sophorose headgroup, sophorose itself being known to have three energetic minima, each associated with a given pair of dihedral angles along the glycosidic, β1,2, bond.[68,69] This was demonstrated for (**5**) by $^{13}$C solid-state NMR arguments.[66] Although the thermodynamic treatment of the self-assembly of symmetrical bolaform amphiphiles generally favors classical morphologies like spherical micelles, vesicles or lamellae, fibrillation is not uncommon.[70–73]

The reason why the micellar phase is more easily observed, and almost quantitative for (**3**), (**4**) and (**1**), while the kinetics of the **M**-to-**F** phase transition seems to be faster for (**2**) and (**5**), certainly deserves further studies. However, it can be observed that (**5**) is the only nonacetylated molecule, thus suggesting that the OH group in 6' and 6'' and its involvement in the hydrogen bonds network could play a significant role. Since this is not unreasonable,[74,75] the situation is remarkable because (**2**) is acetylated. To account for the fibrillation of (**2**), (**2**) should be considered a C18:0, saturated, aliphatic chain, whereas saturated biosurfactants (sophorolipids or cellobioselipids) have shown a stronger tendency to fibrillate than monounsaturated ones.[37,57] In this regard, (**1**), (**3**) and (**4**) are all acetylated and monounsaturated (C18:1). Finally, the role of the acetal bond location, [ω]/[ω-1], in the aliphatic chain is currently unclear, but it does not seem to play a crucial role. This aspect will be discussed towards the end of the manuscript.

*Sophorolipids C18:X (X= 0, 1)* ((**6**)-(**11**) in Figure 5). This class of molecules contains the well-known classical sophorolipids, acetylated and nonacetylated acidic and lactonic C18:1 [ω-1] SL: (**7**), (**8**) and (**9**). (**8**) was studied elsewhere in detail and displayed a micellar phase up to 16 wt%, with the typical tendency to show a sphere-to-wormlike morphological evolution.[76]

This behavior is different from what was observed for bolaform SL and SS discussed above and more consistent with the self-assembly of surfactants in water.[67] Similar results were obtained for (**7**) up to about 4.8 wt%, above which flat structures, in the form of disk-like micelles (**DL**) or bilayers (**Bi**) were recorded (Table 3). This is suggested by the slope in the order of -2 in the corresponding SAXS profiles (6.3 wt%, Figure 6a).

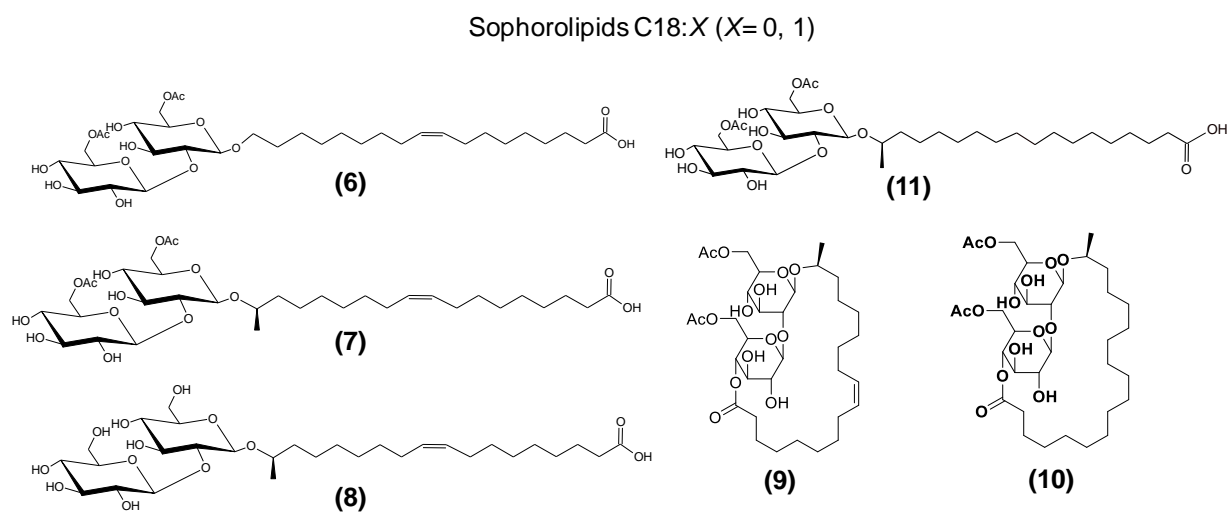

**Figure 5 – Chemical structures of the sophorolipids (SL) studied in this work.**

The apparent broad scattering peak (**BSP**) in the SAXS profiles of (**7**) above about 0.4 wt% (Figure 6a, Figure S 2) could be explained by the presence of repulsive interactions due to partial deprotonation, as discussed elsewhere.[77,78] Comparing the behavior of (**7**) and (**8**) shows that acetylation of 6' and 6'' in sophorose has a partial influence on the morphology of the self-assembled structures, although this is not critical at concentrations below 5 wt%. On the other hand, lactonization, coupled to acetylation, plays a more critical role on self-assembly: (**9**) (lactone) only forms micelles (most likely cylindrical or wormlike, slope > -1/-1.5, Table 4, Figure 6a, Figure S 2) up to 1 wt% and **Bi** structures above (Table 3). The latter is shown by the steeper slope (-2) in SAXS (Table S 1, Figure S 2) and by cryo-TEM images (Figure 4) illustrating a rather flat cross section of the anisotropic objects. This seems to be consistent with the more hydrophobic nature of (**9**) and in agreement with previous data.[79]

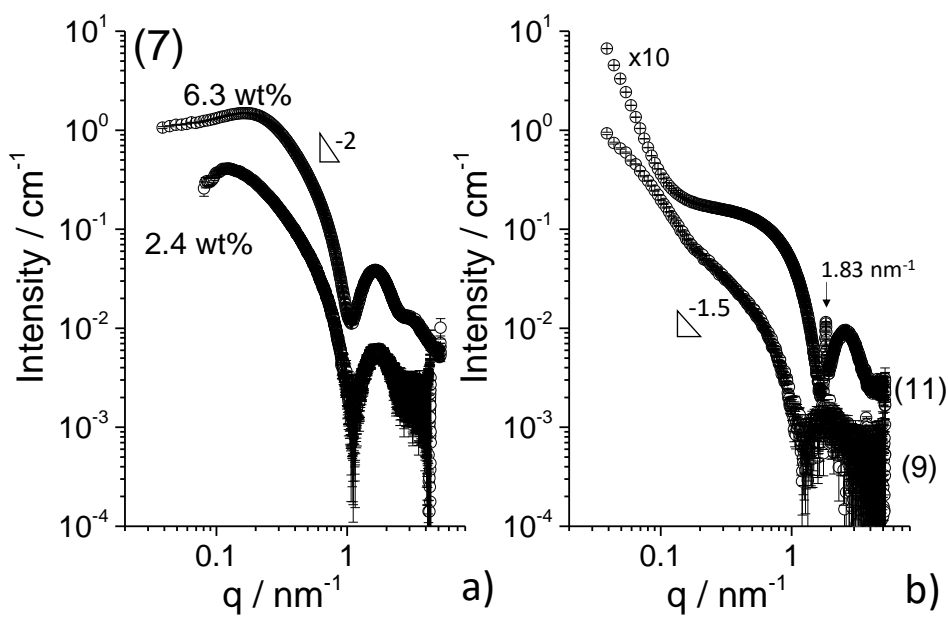

**Figure 6** - SAXS profiles of a) (**7**) and b) (**9**) (1.0 wt%) and (**11**) (2.4 wt%). The multiplication factor is given in the image.

Compound (**6**) differs from (**7**) on the position of the acetal bond. The [ω] position in (**6**) seems to favor a more stable domain of the micellar phase, with a similar sphere-to-cylinder (finally wormlike) transition up to 9 wt% (Table 3, SAXS in Figure S 2). It is possible that flat structures, like disk-like micelles or bilayers as in (**7**) are observed at higher concentrations. Indeed, both (**6**) and (**8**) were studied at concentrations above 20 wt% using a combination of SAXS and polarized optical microscopy. They were shown to form a lamellar phase above 60-70 wt%, with the difference that (**8**) stabilizes multilamellar spherulitic structures.[26]

**Table 3** - Phase behavior of sophorolipids. The $N°$ column refers to the numbers of compounds in Figure 5. M: Micelles, Bi: Bilayers, DL: Disk-like micelles; F: Fibers

| N° | Phase (Concentrations in wt%) | | | | | | | | Comments |
|---|---|---|---|---|---|---|---|---|---|
| | 0.12 | 0.25 | 0.5 | 1.0 | 2.4 | 4.8 | 9.1 | 16.7 | |
| (6) | - | M | M | M | M | M | | | Spheres> Cylinders |
| (7) | M | M | M | M | M | M | DL? or Bi? | | Cylinders >Bilayers + Structure factor |
| (8) | M | M | M | M | M | M | M | M | From Ref. *Green Chem.* **12**, 1564 (2010) |
| (9) | - | M | M + Bi | Bi | Bi | Bi | | | Cylinders/Fibers >Bilayer membranes |
| (10) | | | | | | | | | |
| (11) | - | M + F / F | M + F / F | M + F / F | M + F / F | M + F / F | | | (**M** + **F**) --> **F** with time. Monophase fibrillation. Gel or spherulites depending on cooling rate |

The effect of saturation was observed for compounds (**10**) and (**11**). (**10**) could however not be studied due its lack of solubility at room temperature in the entire concentration range. (**11**), on the other hand, formed a mixture of a micellar and fibrillar phase, the latter characterized by a broad diffraction peak at 1.83 nm$^{-1}$ (Figure 6b, Figure S 2), typically observed in similar systems.[57,58] The micellar phase was stable upon heating above room temperature (here at 80°C using a dry bath) and underwent a micelle-to-fiber transition at room temperature. Hydrogels can occasionally form if the cooling rate is slow, in agreement with the literature on low-molecular weight gelators:[80,81] fibrillar spherulites of (**11**) were observed by optical microscopy (Figure 7) for a cooling rate of 10°C/min from 70°C to 25°C. The fibrillar behavior of (**11**) was not unexpected, as the similar nonacetylated compound SL-C18:0 [ω-1] was described to form crystalline nanoribbons and hydrogels up to 10 wt%.[57,82] The similarities between (**11**) and its nonacetylated counterpart show that acetylation does not play a critical role on the self-assembly of SL.

**Table 4 – Parameters extracted from the SAXS data for sophorolipids. The N° column refers to the numbers of compounds in Figure 5. CS: Clear solution, ColS: Colloidal solution, Rg: Radius of gyration, BSP: Broad Structure Peak, CP: Crystalline Phase, *: Fast heating-cooling (80°C-->25°C), **: Sonication/vortexing.**

| N° | State | pH | Rg (nm) | Slope < 0.08 nm$^{-1}$ | Peaks (nm$^{-1}$) |
|---|---|---|---|---|---|
| (6) | ColS** | 5.0 | 4.4 (0.5%) | 0.0>1.8 (mid-q) (0.1-4.8%) | - |
| (7) | CS | 4.8 | - | 1.0>2.0 (mid-q) (0.1-4.8%) | BSP: 0.27>0.19 (0.1>4.8%) |
| (8) | CS | 4.5 | 1.6 (0.5 wt%) | 0.0>1.6 (mid-q) (0.5-17%) | - |
| (9) | ColS* | 4.2 | - | 1.1>2.0 (0.2>16.7%) | - |
| (10) | Powder | - | - | - | - |
| (11) | ColS* | 6.5 | 1.9 | 3.3 | CP: 1.83; 3.67 |

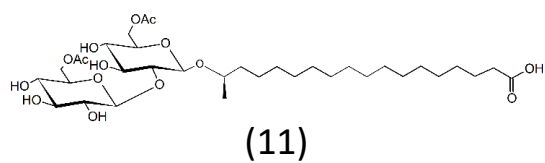

(11)

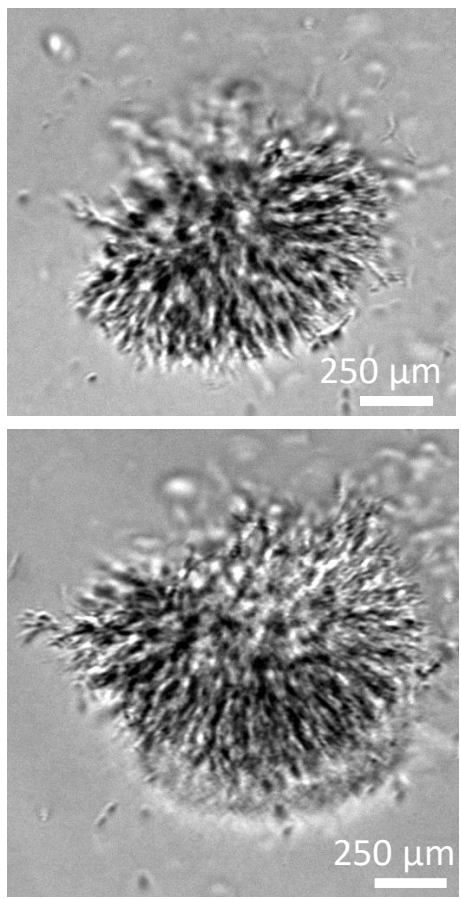

**Figure 7** - Optical microscopy of aqueous solutions of (11) (C(11)= 1 wt%). The sample initially forms a gel at T= 20°C, before observation. The sample is heated to T= 70°C (sol), cooled back rapidly to 20°C (sol), and analyzed.

As a general trend, the least soluble (Table S 1) compounds, like (**9**) and (**11**), appear as turbid colloidal aqueous solutions (**ColS**), and need thermal energy (indicated by the symbol * in Table 4) to improve their homogeneous dispersion. In the case of (**6**), slightly more soluble (Table S 1), mechanical energy like sonication and vortexing seems to be enough. When measurable, the Rg vary from 1.6 nm for (**8**) (estimated from Eq. 1 using the total equatorial radius, $R_{e-c}+T_{e-s}$, of 2.0 nm, measured at pH 5 and reported in Table 7 in Ref. [19]) to 4.4 for (**6**). Broadly-speaking, it was shown before that the micellar diameter of (**8**) corresponds to its molecular length,[83] as expected for bolaform amphiphiles.[64] The larger diameter of (**7**) suggests that acetylation may promote micelles with a larger hydrophobic core. Further analysis of the SAXS data is needed to confirm this point. Lastly, all the SL in this group were studied at mildly acidic pH at which the COOH is mainly protonated. However, this group of compounds is sensitive to pH variation, as shown elsewhere for (**8**).[77,84] Deprotonation at more alkaline pH should not change the phase diagram significantly, and indeed micellization should be enhanced, in a way similar to what is known from previous literature on other biosurfactants.[18,37,77,84,85] This seems to be the case, as suggested by the increase in solubility for all the samples in water (Table S 1). Nevertheless, care must be taken when increasing the pH for lactonic and acetylated compounds, as hydrolysis of the ester bond could occur.

*Glucolipids C18:X (X= 0, 1) [ω-1]* ((**12**), (**13**) in Figure 8). Glucolipids are the mono-glucose analogues of sophorolipids and they are obtained by the fermentation of modified *S. bombicola*. Previous studies demonstrated that glucolipid C18:1 (G-C18:1) forms a stable vesicle phase in water at concentrations up to 0.5 wt%, while glucolipid C18:0 (G-C18:0) forms a hydrated liquid-crystalline lamellar phase up to 10 wt% in water.[37,61,84] These results are valid at pH below 7, while above pH 7 a major micellar phase is observed for both compounds.[37,84] The vesicle and lamellar phases could be obtained by a pH-jump mechanism, from alkaline to acidic, but also by sonication and vortexing, possibly coupled to a heating/cooling cycle at 70°C. Both phases were characterized by classical coupling of SAXS and cryo-TEM, with the exception that the hydrated lamellar phase prepared from G-C18:0 was very defective and its typical SAXS profile only showed a broad peak in the low-q region of the SAXS curves.[61,86] The lamellar phases of G-C18:0 are viscous and, in the case of excess salt ($Na^+$ or $Ca^{2+}$), form lamellar hydrogels at a pH between 5 and 7 and concentrations as low as 1 wt%.[61,86]

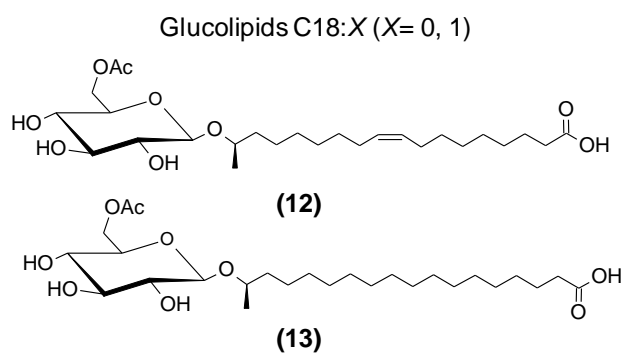

**Figure 8 – Chemical structures of the glucolipids (GL) studied in this work.**

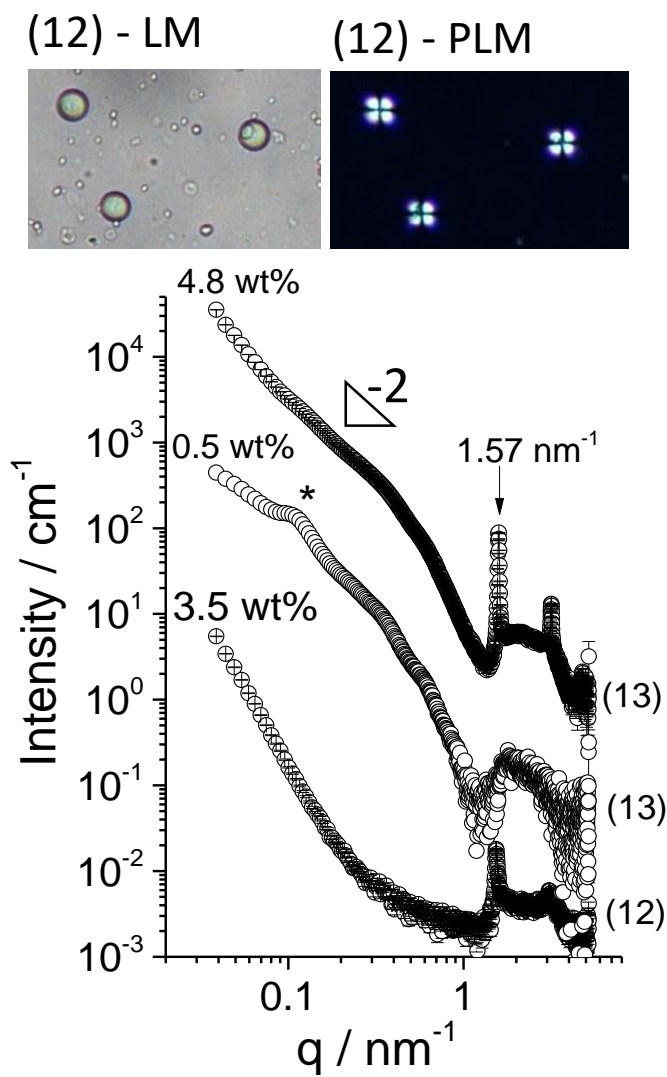

**Figure 9 - SAXS profiles of (12) and (13). The multiplication factor is 100 and 500, respectively for (13) at 0.5 wt% and 4.8 wt%. Light microscopy (LM) and polarized light microscopy (PLM) images (magnification: x20) of (12) collected at 2 wt%.**

Very similar features were observed for the glucolipids (**12**) and (**13**), the acetylated counterparts of G-C18:1 and G-C18:0. Diluted aqueous solutions of (**12**) behaved as stable colloidal suspensions (**ColS**) under diluted conditions (similarly to G-C18:1), while (**12**) was essentially insoluble (lamellar precipitate, **LP**) at higher concentration (Table 5). (**12**) formed a vesicle (**V**) and **V**-**MLV** (multilamellar vesicle) phase mixture below 1 wt%, with a tendency to form **MLV** and a lamellar precipitate (**LP**) above 2 wt%, as summarized in Table 5. SAXS profiles of (**12**) below 1 wt% were typical of **V** and **MLV** ((**12**) in Figure S 2), as also confirmed by light and polarized light microscopy (LM, PLM) images showing spherical objects with a Maltese cross (Figure 9, Figure S 3b). The typical sign of a lamellar precipitate is given by SAXS in Figure 9, characterized by a steep slope in the region of -4 (interface) and diffraction peaks in 1:2 q-ratio.

Table 5 - Phase behavior of glucolipids. The $N^\bullet$ column refers to the numbers of compounds in Figure 8. V: Vesicles, MLV: Multilamellar vesicle; LP: Lamellar Precipitate, L: Lamellar colloidal.

| N° | Phase (Concentrations in wt%) | | | | | | | |
|---|---|---|---|---|---|---|---|---|
| | 0.12 | 0.25 | 0.5 | 1.0 | 2.4 | 4.8 | 9.1 | 16.7 |
| (12) | V | V + MLV | | | | MLV + LP | LP | |
| (13) | Swollen L | | | Swollen L + LP | | | | |

Table 6 – Parameters extracted from the SAXS data for glucolipids [ω-1] C18:$X$ ($X$= 0, 1). The $N^\bullet$ column refers to the numbers of compounds in Figure 8. ColS: Colloidal solution, Rg: Radius of gyration, L: Lamellar colloidal, LP: Lamellar precipitate, BLP: Broad Lamellar Peak, SLP: Sharp Lamellar Peak, *: Fast heating-cooling (80°C-->25°C), **: Sonication/vortexing

| N° | State | pH | Rg (nm) | Slope < 0.08 nm$^{-1}$ | Peaks (nm$^{-1}$) |
|---|---|---|---|---|---|
| (12) | ColS** | 4.9 | - | 1.8>3.7 (0.1>4%) | Low-q BLP: 0.15-0.5 High-q SLP: 1.53, 3.11 |
| (13) | ColS* | 4.3 | - | L: 1.6-2.1 LP: >3 | Low-q BLP: 0.11 High-q SLP: 1.58, 3.17, 4.77 |

On the other hand, aqueous solutions of (**13**) were colloidal under diluted conditions and highly viscous above 1 wt% (Table 6), similar to what was found for G-C18:0. The SAXS profiles of (**13**) (Figure 9 and Figure S 2) were very similar to the SAXS patterns of nonacetylated G-C18:0:[61,86] this includes the broad oscillation above 1 nm$^{-1}$, the -2 slope below 0.8 nm$^{-1}$ and, above all, the broad correlation peak at low-q, as noted by the * (0.12 nm$^{-1}$) in the SAXS pattern of (**13**) at 0.5 wt% in Figure 9. The similarities between the chemical structures of (**13**) and nonacetylated G-C18:1 combined with macroscopical observations and SAXS strongly suggest that (**13**) forms a hydrated, swollen, lamellar phase (**Swollen L**) up to 1 wt% in liquid crystalline state.[61,86] Above 1 wt%, **Swollen L** of (**13**) coexisted with **LP**, as shown by the concomitant presence of a scattering profile with a -2 slope and diffraction peaks (1:2 q-ratio, q1= 1.57 nm$^{-1}$) above 1 nm$^{-1}$ (Figure 9).

All in all, the comparison between (**12**) and (**13**) and their nonacetylated analogues shows that the acetylation of 6' of glucose does not have a critical impact on phase behavior in water, compared, for instance, to the role of unsaturation.

*Sophorolipids/sophorosides C9:0* ((**14**)-(**40**) in Figure 10). This class of biosurfactants is derived from acetylated and nonacetylated sophorolipids, of which the double bond at C9,10 is broken after ozonolysis.[27,87] Acidic (**14**), (**17**), alcoholic (**16**), (**18**), aldehyde (**16**), amine (**19**)-(**24**), (**29**)-(**34**) and ammonium (**25**)-(**28**), (**35**)-(**40**) derivatives are obtained. From a phase behaviour point of view, the end-group of the (**14**)-(**18**) series (acid, alcohol, aldehyde) does not seem to have a major influence (Table 7).

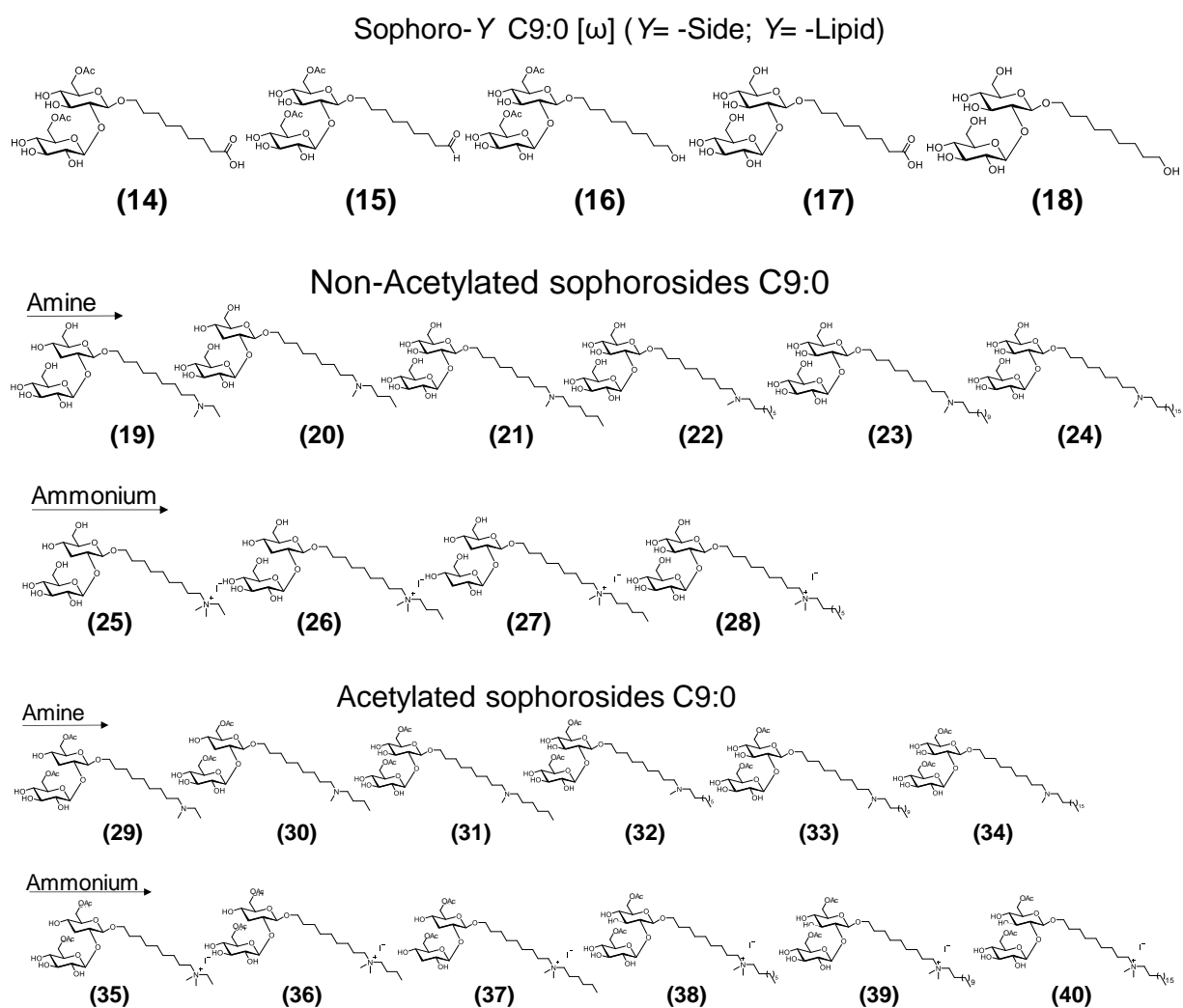

**Figure 10 – Chemical structures of the C9:0 [ω] sophorolipids (SL) and sophorosides (SS) studied in this work.**

(**14**)-(**16**) are highly soluble (CMCs around 0.1 wt%, Table S 1), as they formed clear solutions (**CS**) in the entire range of concentrations (Table 8), while (**17**), (**18**) are turbid colloidal solutions (Table S 1), which need mechanical (** in Table 8) or thermal energy (* in Table 8) to be dispersed. All acetylated samples tend to form micelles (**M**), possibly coexisting with small amounts of large undetermined structures (*LS*) (Table 7), whereas *LS* are characterized by a strong low q scattering signal in SAXS (Figure S 2). The micelles, on the other hand, are small (Rg between 1.2 and 1.5 nm, Table 8), reflecting the smaller size (2.3 nm, using 1 nm for sophorose and 1.3 for a C9 chain according to Eq. 4) of the C9 SL compared to C18 SL. The weak amplitude of the form factor oscillation above 1 nm$^{-1}$ also suggests a poor electron density contrast between water and the micelles.

In the specific case of nonacetylated compounds (**17**), (**18**), the slope at low q in the SAXS profiles being close to -2 (Table 8) suggests that *LS* are composed of platelet aggregates (**P**), as shown for other sophorolipids at basic pH.[88] However, this hypothesis should be verified further. (**18**) is finally insoluble above 9 wt%.

Although the phase behavior between acetylated and nonacetylated C9 samples (**14**)-(**18**) is essentially the same, acetylation seems to play a role in solubility with nonacetylated compounds being more soluble.

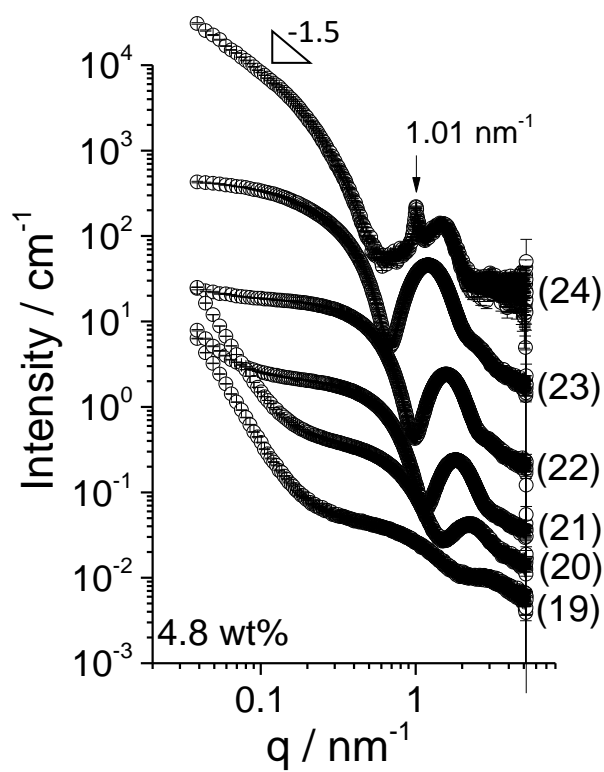

**Figure 11 - SAXS profiles of (12)-(24) at 4.8 wt%. The multiplication factors are 1.5 (20), 8 (21), 50 (22), 500 (23) and 30000 (24).**

The aminyl and ammonium derivatives (**19**)-(**40**) occupied an increasing length (C2 to C18) of the alkyl chain substituent on the nitrogen end-group of C9. The aminyl derivatives could be positively charged at neutral/acidic pH, while the ammonium derivatives were always positively charged. Previous studies on similar compounds (samples 1f-1l in Ref. [89]) showed that ammonium derivatives of C9 and C11 sophrosides tend to self-assemble into micelles, of which the diameter is directly correlated to the length of the alkyl substituent on the nitrogen center. The positive charge on the nitrogen atom is at the origin of repulsive interactions between the micelles, which show a characteristic broad scattering peak in the SAXS profiles.[89]

Similar results were observed for (**19**)-(**40**). Both the nonacetylated (**19**)-(**24**) and acetylated (**29**)-(**34**) aminyl C9 SS tended to form micellar solutions (Table 7), of which Rg increased with the length of the alkyl substituent (Table 8), as shown by the progressive shift of the SAXS profiles towards the low-q (Figure 11 and Figure S 2). Typical Rg values were 1.6, 3.3 and 6.1 nm for the nonacetylated (**19**) (C2), (**21**) (C6) and (**23**) (C12), respectively (Table 8). Another common point between (**19**)-(**24**) and (**29**)-(**34**) was the coexistence between micelles and the large undetermined structures (**LS**) (Table 7), displaying typical strong scattering at low-q, especially at low concentrations and for samples with short alkyl substituents (e.g., (**19**) and (**20**) in Figure 11 and Figure S 2). Although most of these solutions were clear (**CS**), the **LS** could be large enough to form a stable colloidal suspension (**ColS**).

**Table 7 - Phase behavior of C9:0 [ω] sophorolipids and sophorosides. The *N°* column refers to the numbers of the compounds in Figure 10. M: Micellar, *LS*: Large undetermined structures, minor amount, LS: Large undetermined structures, P: Platelet aggregates (possibly), C: Crystalline, INS: Insoluble.**

| N° | Phase (Concentrations in wt%) | | | | | | | | Comments |
|---|---|---|---|---|---|---|---|---|---|
| | 0.12 | 0.25 | 0.5 | 1.0 | 2.4 | 4.8 | 9.1 | 16.7 | |
| (14) | *LS* | | M+*LS* | | | | | | Poor contrast |
| (15) | M | | | | | | M+*LS* | | Poor contrast |
| (16) | M | | | | | | M+*LS* | | Poor contrast |
| (17) | P | | | | | P+M | | | Poor contrast |
| (18) | P | | | | P+M | | INS | | Poor contrast. Insoluble > 9.1 wt% but Soluble upon heating. |
| (19) | *LS* | | M+LS | | | | M+LS + C | | Poor contrast. C = Precipitate after time visible in solution |
| (20) | M | | M+LS | | | | M+LS + C | | Poor contrast. C = Precipitate after time visible in solution |
| (21) | M | | | | M+*LS* | | | | Good contrast |
| (22) | M | | | | | | M+*LS* | | Good contrast |
| (23) | M | | | | | | M+*LS* | | Good contrast |
| (24) | M (cylinders) + C (tapes, Ribbons) | | | | | | | | Proportions vary with time. Monophase fibr./gel in time |
| (25) | - | P | | | | P+M | | | Poor contrast |
| (26) | LS | | | M+LS | | | M+LS + C | | Poor contrast. C = Precipitate after time visible in solution |
| (27) | - | P | | | P+M | | | | Poor contrast |

| (28) | P | P+M | | Poor contrast |
|---|---|---|---|---|
| (29) | M | M+*LS* | | Poor contrast |
| (30) | M | M+LS | | Poor contrast |
| (31) | M+LS | | | Poor contrast. Small precipitate after time. |
| (32) | M | | | Good contrast; Structure factor |
| (33) | M | M+*LS* | | Good contrast; Structure factor |
| (34) | P+M | | P+M + C | Good contrast; Structure factor. C = Precipitate after time visible in solution |

| (35) | M | M+*LS* | | Poor contrast |
|---|---|---|---|---|
| (36) | M+*LS* | | | Poor contrast |
| (37) | M | M+LS | | Poor contrast |
| (38) | M | | M+*LS* | Poor contrast |
| (39) | M | | | Good contrast; Structure factor |
| (40) | M+LS | | M+LS + C | Good contrast; Structure factor. C = Precipitate after time visible in solution |

Homogenization could require heating (* in Table 8) but generally rather vortexing/sonication (** in Table 8). At higher concentration, **LS** became preponderant and insoluble and crystalline (**C**) precipitates were observed. In this regard, the C18 derivative, whether acetylated or not (**34**), (**24**), was characterized by the coexistence of micelles and crystalline ribbons (Figure S 3a). The SAXS profile of (**24**) shows both the features of micelles in solution (low-q slope of -1.5) and crystallinity (peak at 1.01 nm$^{-1}$). Interestingly, crystalline ribbons were mainly observed for sample (**24**) and only rarely for sample (**34**), which seemed to be more characterized by flat, possibly platelet (**P**) structures. Knowledge from other carboxylic (acidic) and aminyl (alkaline) sophorolipids strongly suggests a pH effect.[36]

Previous studies have shown that saturated sophorolipids tend to form crystalline ribbons when they are neutral and a mixture of micelles and platelets when charged.[84,88] This was observed both for carboxylic and aminyl SL,[36] irrespective of the charge, positive or negative, showing that the nature of the functional end-group had relatively little influence. Aminyl derivatives were charged at neutral/acidic pH and uncharged at alkaline pH.[36] Indeed, samples (**24**) and (**34**) had a long saturated alky chain and the pH of their corresponding solutions were alkaline and mildly acidic, respectively. Upon comparison with Ref. [36], it was expected that (**24**) would behave as a neutral saturated SL, with a strong tendency to form crystalline ribbons, and (**34**) to behave as a charged saturated sophorolipid, with a tendency to form a mixture of micelles and platelets. These observations were partially confirmed by the experiments performed with the ammonium derivatives (**25**)-(**28**), (**35**)-(**40**), which tended to form micellar solutions coexisting with **LS** and, possibly, fractal aggregates with possibly platelet (**P**) morphology, although crystalline ribbons were not observed (Table 7). Possible crystalline precipitates occurred at high concentration for the derivative with the longest alkyl chain. Once again, acetylation did not seem to play a critical role in the self-assembly of aminyl and ammonium SS.

**Table 8 – Parameters extracted from the SAXS data for C9:0 [ω] sophorolipids and sophorosides. The $N°$ column refers to the numbers of the compounds in Figure 10. CS: Clear Solution, ColS: Colloidal solution, Rg: Radius of gyration, L: Lamellar colloidal, LP: Lamellar precipitate, \*: Fast heating-cooling (80°C-->25°C), \*\*: Sonication/vortexing, BSP: Broad scattering peak, CP: Crystalline peak.**

| N° | State | pH | Rg (nm) | Slope < 0.08 nm$^{-1}$ | Peaks (nm$^{-1}$) |
|---|---|---|---|---|---|
| (14) | CS | 3.9 | 1.5 | 1.7 | - |
| (15) | CS | 4.6 | 1.4 | 0.9 | - |
| (16) | CS | 6.6 | 1.2 | 0.7 | - |
| (17) | ColS\*\* | 4.4 | - (16 wt%) | 2.3 | - |
| (18) | ColS\* | 8.3 | 1.2 (4.8%) | 2.5-2.2 (0.1-16.7%) | - |
| (19) | ColS\*\* | 10.3 | 1.6 | 2.9 | - |
| (20) | ColS\*\* | 10.4 | 2.3 | 3.2 | - |
| (21) | CS | 10.1 | 3.3 | 1.9 | - |
| (22) | CS | 10.3 | 3.6 | 0.4 | - |
| (23) | ColS\*\* | 9.0 | 6.1 | 0.1 | - |
| (24) | ColS\* (gel>4.8%) | 8.4 | - | 1.6±0.1 | 1.0/1.05 |
| (25) | CS | 6.6 | 1.4 | 2.7 | - |
| (26) | CS | 4.7 | 1.5 | 3.1 | - |
| (27) | CS | 4.5 | 1.5 | 2.9 | - |
| (28) | CS | 5.3 | 1.6 | 2.5 | - |
| (29) | CS | 6.5 | 1.4 | 1.6 | - |
| (30) | ColS\*\* | 7.3 | 1.6 (4.8%) | 3.2 | - |
| (31) | ColS\*\* | 7.3 | - | 3.2 | - |
| (32) | CS | 7.1 | 2.0 | 0.0 | - |
| (33) | CS | 6.7 | - | 1.6 | - |
| (34) | ColS\*\* | 6.5 | - | 2.2-2.7 (0.1-16.7%) | BSP: 0.28>0.43 (0.25>16.7%) CP: 0.55, 1.10, 1.67, 2.2, 2.79 |
| (35) | CS | 3.6 | 1.5 | 0.9 | - |
| (36) | CS | 4.1 | 1.5 | 0.5 | - |
| (37) | ColS\*\* | 6.3 | 1.8 | 2.7>3.2 (0.1-1% ;>1%) | - |
| (38) | CS | 4.7 | 1.5 | 1.8 | - |
| (39) | CS | 4.6 | - | 3.1 | - |
| (40) | ColS\* | 4.3 | - | 3.2 | CP (9.1%): 0.47, 0.95, 1.42 1.89 |

CP (16.7%):
0.99;1.97;2.11;3.92;4.88

Finally, the SAXS profiles of all aminyl and ammonium samples with short (C2-C12) alkyl derivative were characterized by a low amplitude of the first oscillation of the form factor above 1 nm$^{-1}$. SAXS profiles with similar characteristics were reported before for sophorolipids and are generally explained by a poor electron density contrast between the solvent (water) and the micelles.[89] This is most likely due to the strong hydrophilic character of the micelle, with possible water penetration inside the micellar interior. This effect generally disappears with increasing chain length and the formation of micelles with a better defined interface between the hydrophobic core and hydrophilic surface.

**Chemical composition and self-assembly**

This work not only presents the phase diagrams under diluted conditions of a broad set of sophorolipids and sophorosides, but it also helps to better understand the contribution of the structural elements of the SL and SS molecules regarding their self-assembly properties. These are discussed in detail below.

*Acetylation*. Acetyl transferase is the enzyme responsible for acetylation in native SL in position 6' and 6'': the OH group is then replaced by a methyl group. It is then legitimate to question the impact of the acetyl group on the self-assembly of SL and SS. According to the data present, acetyl groups seem to play a role in the physical aspect of the sample. For instance, (**6**), (**7**) and (**12**) are highly viscous samples (Table S 1), difficult to freeze-dry into powders. Nevertheless, despite such a minor difference, the self-assembly properties of acetylated and nonacetylated samples are very similar, with some minor, although potentially significant, differences. For instance, all bolaform SL and SS (**1**) to (**5**) form micellar phases, of which neither the morphology nor the micellar diameter vary as a function of concentration. Although all the samples tended to form a small amount of semi crystalline fibers at long shelf life, only the nonacetylated samples, (**2**) and (**5**), transformed massively into fibrillar hydrogels. The kinetics of fibrillation seemed to be slow, dependent on sample history and probably related to the equilibrium conformation of sophorose.[66] This being said, blocking the 6' and 6'' position with an acetyl group slowed the fibrillation kinetics even more, thus indicating a possible role of these specific OH groups. This is not surprising, as several studies explored the role of sugar conformation on the crystallization of glycosidic amphiphiles. [66,74,75,90,91] Similar considerations regarding the role of the acetyl group are valuable for all the other samples studied in this work. There was no influence of the acetyl groups on the phase behavior of glucolipids C18:1 and C18:0, for instance, forming vesicles and flat lamellae, respectively, both in their acetylated (**12**), (**13**) and nonacetylated form. The micellar structure was preponderant in all C9 SS, irrespective of acetylation and a sphere-to-wormlike transition was observed for both the acetylated and nonacetylated acidic C18:1 SL, which also had similar properties at high concentration (lamellar phase). Similar conclusions on the role of the acetyl group were drawn for other biosurfactants, like MEL-A and porphyrinic derivatives of SL. In particular, only peracetylation of the sugar moiety had a considerable impact on self-assembly properties,[20,92] which were not affected by partial acetylation, as found here.

*[ω] and [ω-1] configuration.* Sophorolipids are generally in [ω-1] configuration, although spurious amounts of [ω] are also synthesized by the wild type *S. bombicola*.[10,31] Recent work has shown that the production of [ω] or [ω-1] configurations is possible,[10,31] with possible consequences on the properties of SL and SS in solution. On the basis of the data collected for this study, the self-assembly properties were not determined by the [ω] or [ω-1] configuration, making this structural parameter of minor importance. The only notable difference between a [ω] or a [ω-1] configuration was the marked presence of disk-like micelles, or bilayer membranes, in the diagram of (**7**) ([ω-1]) compared to (**6**) ([ω]) up to 9 wt% (Table 3). That being said, some differences could be observed at high concentrations. The phase behavior of sophorose-ending (**5**) ([ω] nonacetylated) at concentrations above 60-70 wt% was unclear, while COOH-ending (**6**) ([ω] acetylated) and (**8**) ([ω-1] nonacetylated) displayed the same lamellar phase also at concentrations above 60-70%.[26] This means that the role of the configuration was not after all extremely relevant. Comparison with other works in the literature work is difficult, because, in the present state of the art and to the best of our knowledge, no other studies have specifically explored the role of [ω] or [ω-1] configurations on self-assembly properties.

*End-group.* The interest of SL and SS in comparison to traditional surfactants is the presence of a functional chemical group at the other side of the alkyl chain, opposite sophorose. In this work, several chemical groups were studied, besides the traditional carboxylic acid: amine, ammonium, aldehyde, alcohol, all introduced on the C9 derivatives of SL and SS. According to the present data, none of these end-groups drive critical changes in the self-assembly properties. All the systems formed a micellar phase, often coexisting with large undetermined structures or platelet aggregates. This result was most likely determined by the influence of the C9 chain length, discussed below. This observation does not appear to be greatly dependent on the length of the aliphatic chain, although minor differences were observed. For instance, at room temperature, the C18:1 alcohol derivative of glucolipids (G-C18:1-OH) formed a stable suspension of flat crystalline platelets,[93] while its acidic counterpart (G-C18:1-COOH) precipitated as a lamellar phase in its fully acidic form below pH 4.[37,84] Concerning the aminyl and ammonium derivatives, their behavior seemed to be the same as that of the corresponding acidic counterparts, except that the self-assembly properties were reversed with respect to pH. Micelles and large aggregates formed when the charge was positive on the nitrogen atom (acidic pH for the aminyl derivative) while possible crystallization (sometimes in the form fibers) occurred when the nitrogen atom was uncharged (neutral/alkaline pH for the aminyl derivative).[36] When the end-group contained a $CH_3$, the long chain derivative of SL and SS were insoluble, similar to long-chain (> C14) alkyl polyglucosides. When the end-group was composed of sophorose, as in (**1**) to (**5**), or in other chemical derivatives of SL,[89] micellization was generally improved for a broad set of concentrations (this work and Ref. [45,89]), as predicted by the theory of self-assembly of bolaamphiphiles.[64] Nevertheless, depending on the saturation/unsaturation of the alkyl chain and acetylation degree, fibrillation became a competitive self-assembly process (this work and Ref. [65,66,89]), apparently depending on the conformation of sophorose itself.[66] Sophorose known to have at least three minima of conformational energy.[68,69] Since crystallization depending on the conformation of sugars is not uncommon,[90,91] it is still unclear, given the present state of the art, how to control this process for sophorolipids.[66]

*Lactonization*. Lactonization is due to the action of the *S. bombicola* lactone esterase (*sble*) enzyme in *S. bombicola* and the lactone SL congener is an important compound in the final raw mixture.[10,31] Deletion or overexpression of *sble* leads to the production of a lactone-free or lactone-rich SL, respectively.[10,47,94] In this work, two lactones (**9**), (**10**) were tested, both acetylated, and their behavior depended on the unsaturation of the lipid chain. As a general trend, lactone SL were more insoluble than their acidic counterparts, as reported elsewhere.[79] This was also verified here. In particular, the C18:0 derivative (**10**) was totally insoluble at room temperature (Table S 1) and could not be studied, while the C18:1 derivative (**9**) was mildly soluble (< 0.1 wt%, Table S 1), with a strong tendency to self-assemble into two-dimensional membranes at concentration above 1 wt%, in agreement with Ref. [79]. Compared the acidic SL, which tends more to form micelles, lactonization drove critical variations in the self-assembly properties.

*Chain length*. The length of the alkyl chain is a parameter that has a strong impact in the self-assembly properties of surfactants.[95] As predicted by the thermodynamic theory of self-assembly of amphiphiles, for an equivalent surface area of the hydrophilic headgroup, shorter chains improve solubility and the formation of spherical micelles.[39] However, although it would be interesting to study, the controlled variation of chain length in sophorolipids is not straightforward[27] and in this work only the C18 and C9 derivatives could be studied. In this framework, the comparison between (**8**) (C18) and (**17**) (C9) was instructive. First of all, (**17**) was much more soluble (S> 2 wt% and CMC> 5 wt% at pH 4, Table S 1) and the corresponding SAXS profiles showed that there was no clear-cut scattering signal of micelles. At 16 wt%, (**17**) formed micelles, of which the Rg was in the region of 1.2 nm, while the typical Rg of (**8**) could be estimated from previous work to be in the region of 2.0 nm (at pH 5, see sample 1 or sample 2 in Table 7 in Ref. [19]). According to Eq. 4 (and using the value of 1 nm for the length of a typical disaccharide,[62] like sophorose), (**8**) was expected to have a length of 3.4 nm, while (**17**) of 2.3 nm. These considerations showed that: 1) the micellar radius generally corresponds to half the length of the SL molecules, as predicted for micelles composed of bolaform amphiphiles;[64] 2) the micellar diameter is directly related to the length of the fatty acid. Similar results were found on the series of amine and ammonium C9 derivatives, of which the micellar radius progressively increases with the length of the alkyl substituent on the nitrogen atom.[89] These results were expected and consistent with the broad literature on surfactant self-assembly in water.[96]

*pH*. pH is a parameter having critical effects on the self-assembly of sophorolipids, as largely discussed elsewhere (section 3.6.2 and Table 10 in Ref. [19]). Often, the deprotonated and acidic forms have very different self-assembly properties in water. For instance, the nonacetylated acidic C18:0 SL self-assembles into micelles at high pH (deprotonated) and into twisted ribbons at low pH (protonated).[57] Similar results were found for the aminyl derivative of C18:0 SL, but at inversed pH.[36] In this work, pH was not a parameter of study and all the sample solutions were studied at the equilibrium pH, reported in the data tables. Use of buffer was not an option for several reasons, mainly to avoid potential salt effects: at the present state of the art on biosurfactant science, pH effects are still better understood[19] than salt effects, the latter being much less studied so far.[78,97–99] Despite the lack of a dedicated investigation on the effect of pH, some typical features of pH effects could be observed. For instance, the presence of semicrystalline ribbons for (**24**), and resulting consequent gelation, was most likely related to the slightly alkaline pH of the corresponding aqueous solution. (**24**) is an aminyl derivative, neutral at alkaline pH and with a long, saturated, chain length. Its behavior is analogous to the aminyl C18:0 SL, forming ribbons at alkaline pH.[36] On the other hand, (**34**) is the acetylated counterpart of (**24**) and its stable micellar phase is most likely correlated to a positive change on the nitrogen, explained by the slightly acidic pH of the corresponding solution. The broad scattering peak at about 0.6 nm$^{-1}$ observed for (**33**) and (**34**) (Figure S 2) are also characteristics of repulsive interactions between charged micelles. A similar peak was also observed for the ammonium derivatives (**39**) and (**40**) as well as for a broad set of SL and derivatives of SL.

*(Un)saturation.* Sophorolipids from wild type *S. bombicola* are monounsaturated.[10,31] However, the use of saturated or polyunsaturated fatty acids as feedstock, or catalytical hydrogenation as post-treatment, can generate poly- or saturated SL.[57,100–102] Previous studies have shown the critical impact of the saturation/unsaturation of SL on their self-assembly properties in water.[57,100,101] Polyunsaturated SL tend to form vesicles[100] while saturated SL assemble into fibers,[57] the most prominent example being the nonacetylated C18:0 SL. Similar conclusions can be drawn from the present study. (**11**), the acetylated counterpart of C18:0 SL, fibrillated. Hydrogelation was indeed possible, most likely depending on the cooling strategy, similar to other low molecular weight gelators. The saturated lactonic SL, (**10**), was insoluble and could not even be studied at low concentrations. The bolaform acetylated SS (**2**) rapidly fibrillated and formed a hydrogel, differently from its monounsaturated counterpart, (**3**), of which the micellar phase seemed to be more kinetically stable. Saturation confirmed an important structural change, largely affecting the phase behavior of SL and SS.

*Number of sugar units.* By definition, native sophorolipids contain sophorose, a β1,2 D-glucose disaccharide.[10,31] However, either by enzymatic conversion[103] or by fermentation using modified *S. bombicola* strains,[104,105] it is possible to selectively produce single-glucose C18:1 lipids, G-C18:1 (**12**), which can be converted into C18:0 glucose lipids, G-C18:0 (**13**). Single glucose lipids can be either acetylated (**12**), (**13**) or nonacetylated, the latter studied elsewhere.[37,61,84] Reducing the number of sugar groups has a critical impact on the self-assembly properties. Nonacetylated G-C18:1 and G-C18:0 were largely studied at acidic pH and reported to form vesicle and hydrated lamellar structures, respectively.[37,61,84] They are to be compared with micelles and fibers, respectively obtained with C18:1 and C18:0 sophorolipids under the same physicochemical conditions.[57,77,79] In the present work, (**12**) and (**13**) were acetylated but behaved similarly to their nonacetylated counterparts, thus confirming that reducing the number of glucose in the headgroup is critical for self-assembly, differently from the acetylation of 6'.

**General considerations on structure-properties**

According to the thermodynamic theory of amphiphile self-assembly, the chemical nature and geometry of an amphiphile drive its self-assembly properties.[39,106] Modifications of the chemical structure can therefore have a critical impact on self-assembly and phase behavior. Does this assumption hold for sophorolipids, knowing that the classical theory has been demonstrated to have limits for amphiphiles with more complex structures[40,43,67]? Figure 12 shows some general considerations associating specific modifications of the sophorolipid structure and the possible impact on its self-assembly in solution. It is now possible to better understand the structure-properties relationship of sophorolipids, discussed below from the least to the most critical.

*Not critical*. Acetylation and the position of the acetal bond [ω] or [ω-1] do not affect phase behavior. Acetylation blocks the H-bonding on the 6' and 6'' positions and it introduces a more hydrophobic group, but these changes are not sufficient to considerably change the collective self-assembly of SL and SS in water. Previous work on SL derivatives and mannosylerythritol lipids showed that full acetylation is necessary to substantially change the solution behavior of biosurfactants ([20,92] and comments in refer to 3.5.5 in Ref. [19]). Similarly, the acetal bond on the [ω] or [ω-1] does not change the effective chain length sufficiently to modify the self-assembly. It could be argued that the [ω-1] introduces a chiral center. However, the self-assembly of sophorolipids does not seem to be driven by chirality. For instance, twisted fibrous ribbons were observed both with [ω] (e.g., (**2**), (**5**) and C16:0 derivative of SL[58]) and [ω-1] (e.g., (**11**), C18:0 derivative of SL)[57] derivatives.

## Sophorolipid and glucolipid bioamphiphiles: influence of chemical groups on self-assembly

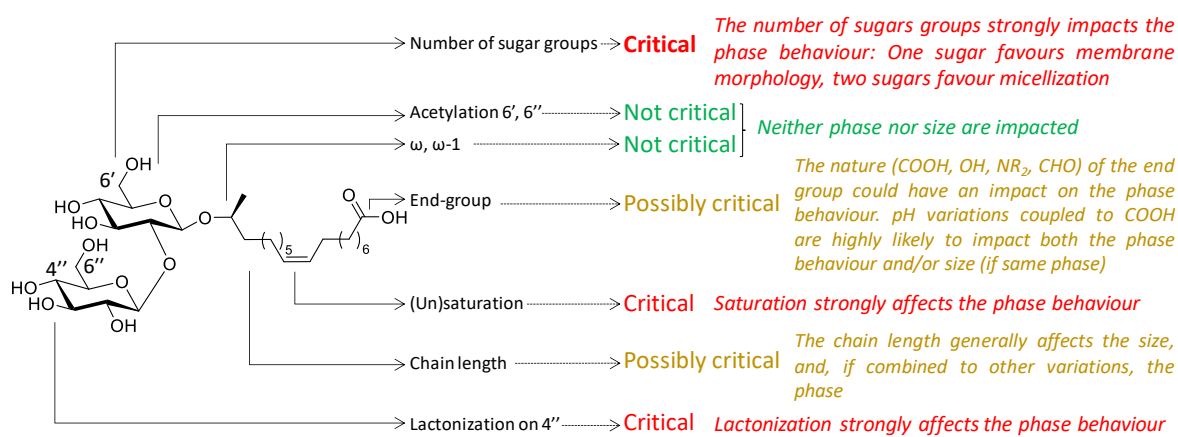

**Figure 12** – Typical chemical groups illustrated on a standard acidic C18:1 sophorolipid and their respective impact on the self-assembly behavior.

*Possibly critical*. The modification of the end-group and chain length may have critical impacts on self-assembly properties, all depending on the nature of the chemical group and length of the alkyl chain. Concerning the chain length, the present work does not show a critical impact of switching between C18 and C9. However, it must be said that the C18 SL is monounsaturated and the C9 derivative should probably be compared to the C18:0 SL. In this regard, reducing the length from C18 to C9 strongly improves solubility and drives self-assembly from fibrillar (C18, at acidic pH) to micellar (C9). In this regard, the impact is critical. This was indeed expected and in agreement with general knowledge of amphiphile self-assembly. At the present state of the art, however, sophorolipids with moderate variation of the fatty acid length are not yet available, especially with a single unsaturation. Further work must therefore be performed at both the level of biosynthesis and physicochemical characterization to better understand the effect of chain length on phase behavior.

Modifications of the end-group are common in SL and the effects cannot be predicted with the theory of self-assembly, which was developed for standard head-tail amphiphiles[39,106] and only partially adapted to simple bolaamphiphiles.[64] Previous studies have explored the possibility of introducing aminyl,[36] alkyl,[107–109] porphyrinyl,[92,110,111] alkinyl,[36] nitrodopamine[112] groups on C18 SL, to mention but a few. Within the framework of this work, alcohol, aldehyde, aminyl and ammonium were tested on the C9 derivative of SL. These modifications did not show a critical evolution of the self-assembly properties. All the samples stabilized micellar phases, more or less coexisting with poorly defined larger structures in a difficult-to-control amount. Indeed, reversing the charge from negative (carboxylate) to positive (ammonium) simply reverses the pH at which specific structures are observed, but not necessarily the type of structure. This result was also shown elsewhere.[36] Nevertheless, the insertion of complex end-groups, like porphiryl[92,110,111] or alkyl,[107–109] or even methyl, can drastically change self-assembly properties. As a general trend, it could be possible to use the rule of thumb according to which, if the end-group is charged or neutral but hydrophilic, the resulting behavior of the derived SL is most likely similar to the classical acidic SL shown in Figure 12, with the possible reversibility of the pH effect. On the other hand, if the end-group is apolar, or a more complex chemical function (e.g., nitrodopamine)[112], one should expect a strong variation in the phase behavior. These modifications must be studied *ex novo*.

*Critical*. Modifications of the number of sugars in the headgroup, saturation, lactonization and state of charge of the end-group are all critical with respect to the self-assembly and phase behavior of SL. All of them play an important role on the geometry of SL and enhance non-specific interactions (e.g., hydrophobic effect, electrostatic or steric repulsion). The change in terms of self-assembly can be partially understood, although not necessarily predicted, with the thermodynamic theory of amphiphile self-assembly.

Sophorose and glucose have quite different surface areas, 70 Å$^2$ [113] and 40-50 Å$^2$ [103,114,115], respectively, and both sophoro and glucolipids have a C18 tail. These values can be compared to the headgroup area of phospholipids (50-60 Å$^2$),[39] which also have an 18-carbon aliphatic chain. One can then expect single glucose lipids to stabilize less curved geometries, vesicles or lamellae, compared to micelles. This is exactly what happens when the end-group is neutral, meaning that the self-assembly of neutral monounsaturated (C18:1) gluco and sophorolipids could be described using the thermodynamic theory of self-assembly. On the other hand, this approach fails when the end group is charged (positive or negative), most likely because the charge introduces an additional hydrophilic headgroup with a large surface area, due to repulsive electrostatic interactions, as classically found for cationic and anionic surfactants. The resulting assembled structures are generally rich in micelles, independently of the type of headgroup (refer to the paragraph above about the effect of pH).

Concerning the fatty acid tail, its lactonization has a tendency to drive the assembly towards membrane structures. If taking the packing parameter approach, it is not unreasonable to consider that the effective length of the fatty chain in a lactone configuration is at least half that of the fully extended chain, while the effectively occupied volume is double. The shorter length and larger volume of the hydrophobic tail increase the packing parameter, thus resulting in more flat structures. Experimentally, lactone SL tend to form flat membranes, disk-like micelles or vesicles. On the other hand, eliminating the *cis* double bond from sophorolipids yields saturated lipids, with a straightforward increase in the melting temperature, as found experimentally for oleic (13°C) and stearic acid (69°C). Under these conditions, the thermodynamic theory of self-assembly is no longer valid, as it assumes a liquid-like aliphatic core.[39] In the absence of charges, the nature of the resulting structures differs from the C18:1 SL, but it strongly depends on the number of sugar headgroups. Indeed, crystalline fibers, often taking the form of twisted ribbons, are observed for saturated sophorolipids and sophorosides. The ribbons are crystalline and the twist, as explained for similar systems in the literature, seems to be due to the steric hindrance of the sophorose headgroup rather than to a chiral effect (3.6.1 in Ref. [19]). On the other hand, single glucose saturated lipids tend to form rigid interdigitated membranes. Although the glycosidic, hydrophilic nature of the headgroup guarantees the hydration of the fibers and membranes, their colloidal dispersion in solution, with possibility to form hydrogels, actually depends either on the method of synthesis or on the ratio between the neutral and charged form of the lipid. In this regard, the overall collective behavior of saturated sophorolipids and glucolipids cannot be predicted beforehand.

**Conclusions**

The backbone of a sophorolipid molecule is constituted by a sophorose (glucose β1,2) headgroup covalently linked to a C18:1 fatty acid through an acetal bond, where the carboxylic acid is opposite the sophorose. In view to understanding the impact of chemical groups on the self-assembly properties of SLs, 40 different chemical variations were studied in water at room temperature under diluted and semi-diluted (< 20 wt%) conditions.

The replacement of the carboxylic acid by another sophorose group provides a family of compounds named sophorosides (**1**)-(**5**), of which the aggregation type is essentially micellar with the ability to assemble into fibers in time for (**2**) and (**5**). Fluffy fibrous aggregates can occasionally be found for (**1**), (**3**) and (**4**).

The effect of acetylation, saturation, lactonization and the position of the acetal bond are studies on compounds (**6**)-(**10**). The acetal bond position (ω or ω-1) and acetylation on 6' and 6'', samples (**6**), (**7**), (**8**) do not have a critical impact on self-assembly. The micellar phase of the corresponding SL derivatives is mostly unchanged. Lactonization promotes bilayer structures (**9**) while saturation (**11**) promotes fibrillation. The combination between lactonization and saturation generates an insoluble compound (**10**).

Eliminating a glucose molecule from the headgroup (**12**)-(**13**) results in a lower surface area at the water-micellar core interface, thus promoting flat structures, like vesicles or lamellae, while reducing the chain length to a C9 (**14**)-(**18**) improves solubility and the tendency to form small micellar aggregates. Introducing an amine or ammonium group at the end of a C9 backbone, with or without acetylation on the sophorose headgroup (**19**), (**25**), (**29**), (**35**) does not have a major impact on self-assembly: poorly contrasted small micellar aggregates dominate. Micellization is again promoted by increasing the length of the acyl chain substitution on the nitrogen group (**20**)-(**24**), (**26**)-(**28**), (**30**)-(**34**) and (**36**)-(**40**).

The broad set of data collected in this work on 40 different compounds illustrate a preliminary structure-properties relationship never attempted before on sophorolipids. It was observed that the impact of the acetal bond position (ω or ω-1) and acetylation on 6' and 6'' on self-assembly are limited, while the chain length and the chemical nature of the end-group have a potentially critical impact on the self-assembly of SL and SS. The number of sugar units in the headgroup, the (un)saturation in the alkyl chain, lactonization of the carbonyl and pH (for the acidic, COOH end-group, molecules) have a systematic critical impact on phase behavior.

When considering the field of surfactant science, SL seems to fall in the domain of exceptions,[43] for which a general predictive self-assembly theory does not yet exist. The results generated in this work should contribute to more experimental and theoretical work in order to eventually fully understand and control the properties in solution of this important class of bioamphiphiles.

**Acknowledgements**

ESRF (proposal number MX2311) is gratefully acknowledged for its financial support. Leopold Mottet (Flamac, Gent, Belgium) is warmly acknowledged for his assistance in the surface tension and precipitation experiments. VLAIO - Flanders innovation & entrepreneurship – is acknowledged for its financial support to the APPLISURF Project.

**Open data**

All SAXS profiles are shown in the Supporting Information while all *.dat files are grouped, zipped and freely available at https://doi.org/10.5281/zenodo.7677356

# Supporting information

**Aqueous self-assembly of a wide range of sophorolipid and glucolipid microbial bioamphiphiles (biosurfactants): considerations about the structure-properties relationship**

Niki Baccile[a,*] Alexandre Poirier,[a] Patrick Le Griel,[a] Petra Pernot,[b] Melike Pala,[c] Sophie Roelants,[d,e] Wim Soetaert[d,e] Christian V. Stevens[c]

[a] Sorbonne Université, Centre National de la Recherche Scientifique, Laboratoire de Chimie de la Matière Condensée de Paris, LCMCP, F-75005 Paris, France

[b] ESRF – The European Synchrotron, CS40220, 38043 Grenoble, France

[c] Ghent University, Faculty of Bioscience Engineering, Department of Green Chemistry and Technology, Coupure Links 653, B-9000 Gent, Belgium

[d] Centre for Industrial Biotechnology and Biocatalysis (InBio.be), Department of Biotechnology, Faculty of Bioscience Engineering, Ghent University, Coupure Links 653, 9000 Ghent, Belgium

[e] Bio Base Europe Pilot Plant, Rodenhuizenkaai 1, 9042 Ghent, Belgium

\* Corresponding author:
Dr. Niki Baccile
E-mail address: niki.baccile@sorbonne-universite.fr
Phone: +33 1 44 27 56 77

## Molecular structures, synthesis and analysis

Bolaform Sophoro-*Y* (*Y*= -Side; *Y*= -Lipid)

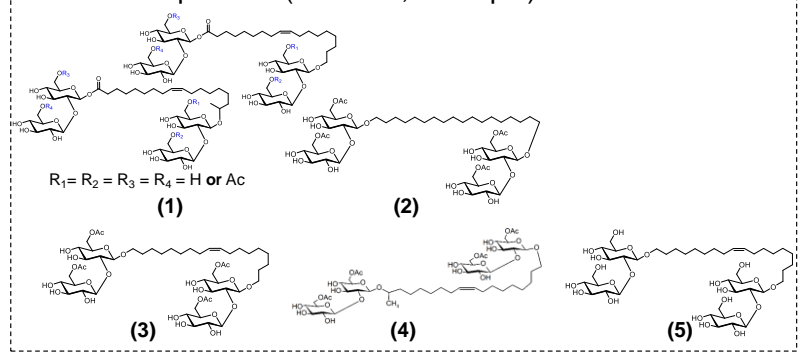

$R_1 = R_2 = R_3 = R_4 = $ H **or** Ac

Sophorolipids C18:*X* (*X*= 0, 1)

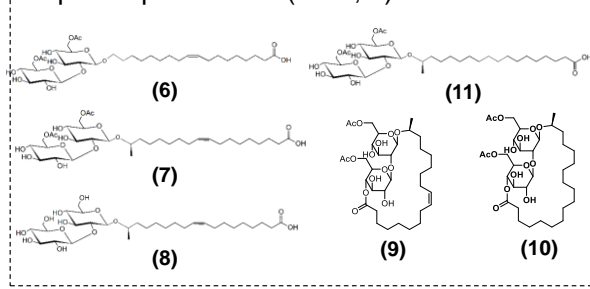

Glucolipids C18:*X* (*X*= 0, 1)

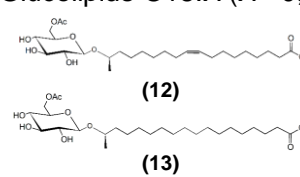

Sophoro-*Y* C9:0 [ω] (*Y*= -Side; *Y*= -Lipid)

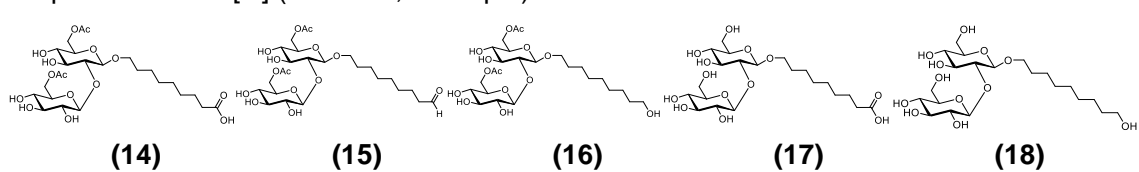

Non-Acetylated sophorosides C9:0

Amine

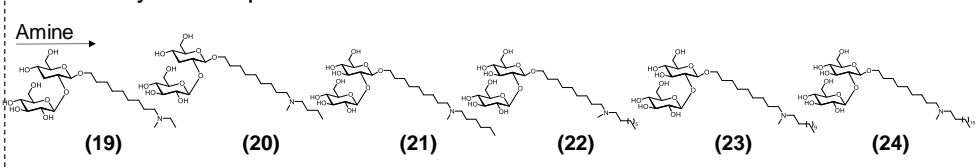

Ammonium

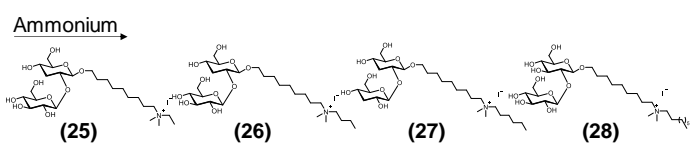

Acetylated sophorosides C9:0

Amine

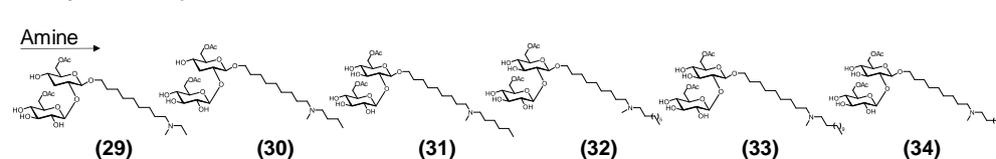

Ammonium

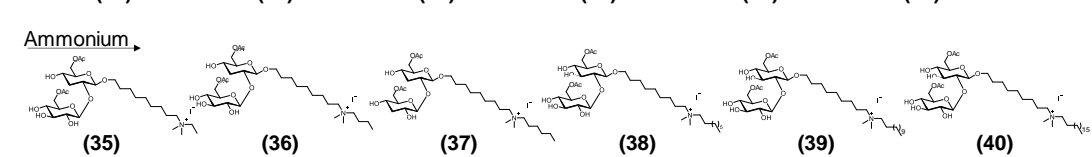

**Solution NMR analysis for compounds (6), (29), (30), (32), (33), (34), (35), (36), (37), (38), (39), (40)**

**Acetylated acidic sophorolipid ω C18:1 (6)**

*Determined from a mixture (~50/50) of non-acetylayed and acetylated C18:0 ω sophorolipid acid together with some congeners. Only acetylated C18:0 ω sophorolipid acid is described. Isoproponal and acetic acid peaks are also present.*

**$^1$H NMR** (400 MHz, DMSO): $\delta_H$ 1.20-1.47 (26H, m, 13xCH$_2$(CH$_2$)$_2$), 1.52-1.72 (4H, m, CH$_2$CH$_2$(C=O)OH, CHOCH$_2$CH$_2$), 1.80-1.92 (4H, m, 2xCH$_2$CH=CH) 1.92 (6H, s, 2xCH$_3$C=O), 2.29 (2H, t, $J$ = 7.6 Hz, CH$_2$(C=O)OH), 2.99-3.10 (1H, m, C$^{2''}$H), (3.08-3.22 (2H, m, 2xCHOC) (3.10-3.19 (1H, m, C$^{3''}$H), (3.19-3.27 (1H, m, C$^{2'}$H), 3.26-3.35 (2H, m, 2xCHOC), 3.36-3.44 (2H, m, C$^{3'}$H, CHOCH$_a$H$_b$CH$_2$), 3.96-4.08 (1H, m, CHOCH$_a$H$_b$CH$_2$), 4.12-4.22 (2H, m, CH$_a$H$_b$OAc), 4.24 (1H, m, C$^{1''}$H), 4.34-4.46 (3H, m, 2xCH$_a$H$_b$OAc, C$^{1'}$H), 5.19 (2H, m, HC=CH). **$^{13}$C NMR** (100 MHz, DMSO): $\delta_C$ 20.7 (2xCH$_3$OAc), 24.8 (CH$_2$CH$_2$(C=O)OH), 26.9 (2xCH$_2$CH=CH), 28.7-29.9 (12xCH$_2$(CH$_2$)$_2$), CHOCH$_2$CH$_2$), 34.0 (CH$_2$(C=O)OH), 63.7 (2xCH$_2$OAc), 69.2 (CHOCH$_2$CH$_2$), 69.8 (CHOC), 69.9 (CHOC), 73.3 (CHOC), 73.9 (CHOC), 74.6 (C$^{2''}$H), 75.9 (C$^{3''}$H), 76.1 (C$^{3'}$H), 82.8 (C$^{2'}$H), 101.4 (C$^{1'}$H), 104.2 (C$^{1''}$H), 129.7 (HC=CH), 171.1 (CH$_3$C=O), 171.2 (CH$_3$C=O), 175.5 (HC=O).

**General procedure for synthesis of acetylated sophoroside amines (29), (30), (32), (33), (34).**

In a 300 mL Parr reaction bottle, acetylated soprohoside aldehyde was dissolved in 50 mL ethanol and 1 eq secondary amine was added subsequently. The reaction mixture was stirred for 30 min at room temperature. To was added 5 wt% Pd/C under N$_2$ atmosphere and molecular sieves (3 Å) and the reaction mixture was stirred overnight (18h) under 5 barg H$_2$. The reaction mixture was filtered over celite and the residue was washed with ethylacetate. The filtrare was concentrated under reduced pressure. The acetylated sophorolipid amines were obtained in high purity without further purification.

**$N$-ethyl-$N$-methyl-9-[(6',6''-di-$O$-acetyl-2'-$O$-$\beta$-$D$-glucopyranosyl-$\beta$-$D$-glucopyranosyl)oxy]nonan-1-amine (29):** (7.49 g, 81%), beige powder.

**$^1$H NMR** (400 MHz, DMSO-d$_6$): $\delta_H$ 0.95 (3H, t, $J$ = 7.1 Hz, CH$_2$CH$_3$), 1.13-1.33 (10H, m, 5xCH$_2$(CH$_2$)$_2$), 1.33-1.44 (2H, m, CH$_2$CH$_2$N), 1.45-1.56 (2H, m, CHOCH$_2$CH$_2$), 2.01 (3H, s,

CH₃C=O), 2.02 (3H, s, CH₃C=O), 2.09 (3H, s, CH₃N), 2.23 (2H, t, *J* = 7.5 Hz, CH₂CH₂N), 2.26-2.36 (2H, m, CH₂CH₂N), 3.02 (1H, dxd, *J* = 8.2, 8.2 Hz, C²''H), 3.05-3.19 (3H, m, 2xCHOC, C³''H), 3.22 (1H, dxd, *J* = 8.6, 8.6 Hz, C²'H), 3.28-3.43 (3H, m, 2xCHOC, C³'H), 3.42-3.50 (1H, m, CHOCH$_a$H$_b$CH₂), 3.71-3.80 (1H, m, CHOCH$_a$H$_b$CH₂), 3.98-4.09 (2H, m, 2xCH$_a$H$_b$OAc), 4.22 (1H, d, *J* = 13.1 Hz, CH$_a$H$_b$OAc), 4.26 (1H, d, *J* = 12.2 Hz, CH$_a$H$_b$OAc), 4.32 (1H, d, *J* = 7.5 Hz, C¹'H), 4.41 (1H, d, *J* = 7.6 Hz, C¹''H). **¹³C NMR** (100 MHz, DMSO-d₆): δ$_C$ 12.6 (CH₂CH₃), 21.1 (2xCH₃C=O), 25.8 (CH₂(CH₂)₂), 27.2 (CH₂(CH₂)₂), 27.4 (CH₂CH₂N), 29.3 (CH₂(CH₂)₂), 29.4 (CH₂(CH₂)₂), 29.4 (CH₂CH₂N), 29.7 (CHOCH₂CH₂), 41.7 (CH₃N), 51.3 (CH₃CH₂N), 57.1 (CH₂CH₂N), 64.0 (CH₂OAc), 64.2 (CH₂OAc), 69.1 (CHOCH₂CH₂), 70.1 (CHOC), 70.3 (CHOC), 73.8 (CHOC), 74.3 (CHOC), 75.2 (C²''H), 76.26 (C³'H), 76.30 (C³''H), 83.0 (C²'H), 101.5 (C¹'H), 104.7 (C¹''H), 170.68 (CH₃C=O), 170.71 (CH₃C=O).

***N*-butyl-*N*-methyl-9-[(6',6''-di-*O*-acetyl-2'-*O*-β-D-glucopyranosyl-β-D-glucopyranosyl)oxy]nonan-1-amine (30):** (6.90 g, 76%), off-white powder.

**¹H NMR** (400 MHz, DMSO-d₆): δ$_H$ 0.87 (3H, t, *J* = 7.3 Hz, CH₂CH₃), 1.16-1.33 (12H, m, 5xCH₂(CH₂)₂, CH₂CH₃), 1.33-1.56 (6H, m, 2xCH₂CH₂N, CHOCH₂CH₂), 2.01 (3H, s, CH₃C=O), 2.02 (3H, s, CH₃C=O), 2.09 (3H, s, CH₃N), 2.19-2.27 (4H, m, 2xCH₂CH₂N), 3.01 (1H, dxd, *J* = 8.5, 8.5 Hz, C²''H), 3.05-3.19 (3H, m, 2xCHOC, C³''H), 3.23 (1H, dxd, *J* = 8.6, 8.6 Hz, C²'H), 3.28-3.43 (3H, m, 2xCHOC, C³'H), 3.42-3.50 (1H, m, CHOCH$_a$H$_b$CH₂), 3.71-3.80 (1H, m, CHOCH$_a$H$_b$CH₂), 3.94-4.07, 3.28-3.49 (4H, m, 2xCHOC, C³'H, CHOCH$_a$H$_b$CH₂), 3.71-3.80 (1H, m, CHOCH$_a$H$_b$CH₂), 3.96-4.07 (2H, m, 2xCH$_a$H$_b$OAc), 4.18-4.29 (2H, 2xCH$_a$H$_b$OAc), 4.33 (1H, d, *J* = 7.7 Hz, C¹'H), 4.42 (1H, d, *J* = 7.4 Hz, C¹''H). **¹³C NMR** (100 MHz, DMSO-d₆): δ$_C$ 14.4 (2xCH₂CH₃), 20.5 (2xCH₂CH₃), 21.1 (2xCH₃C=O), 25.8 (CH₂(CH₂)₂), 25.9 (CH₂(CH₂)₂), 27.2 (CH₂CH₂CH₂N), 29.3 (CH₂(CH₂)₂), 29.4 (2xCH₂(CH₂)₂), 29.7 (CHOCH₂CH₂), 42.7 (CH₃N), 57.3 (CH₂CH₂CH₂N), 57.6 (2xCH₂CH₂CH₃), 64.0 (CH₂OAc), 64.2 (CH₂OAc), 69.1 (CHOCH₂CH₂), 70.1 (CHOC), 70.2 (CHOC), 73.8 (CHOC), 74.3 (CHOC), 75.2 (C²''H), 76.27 (C³'H), 76.34 (C³''H), 83.0 (C²'H), 101.6 (C¹'H), 104.7 (C¹''H), 170.65 (CH₃C=O), 170.70 (CH₃C=O).

***N*-hexyl-*N*-methyl-9-[(6',6''-di-*O*-acetyl-2'-*O*-β-D-glucopyranosyl-β-D-glucopyranosyl)oxy]nonan-1-amine (31):** (4.04 g, 68%), beige powder.

**¹H NMR** (400 MHz, DMSO-d₆): δ$_H$ δ$_H$ 0.86 (3H, t, *J* = 6.2 Hz, CH₂CH₃), 1.13-1.32 (16H, m, 7xCH₂(CH₂)₂, CH₂CH₃), 1.32-1.43 (4H, m, 2xCH₂CH₂N), 1.43-1.58 (2H, m,

CHOCH$_2$CH$_2$), 2.00 (3H, s, CH$_3$C=O), 2.02 (3H, s, CH$_3$C=O), 2.08 (3H, s, CH$_3$N), 2.23 (4H, t, $J$ = 7.2 Hz, 2xCH$_2$CH$_2$N), 3.01 (1H, dxd, $J$ = 8.2, 8.2 Hz, C$^{2''}$H), 3.04-3.18 (2H, m, 2xCHOC), 3.13-3.24 (1H, m, C$^{3''}$H), 3.22 (1H, dxd, $J$ = 8.2, 8.6 Hz, C$^{2'}$H), 3.28-3.35 (1H, m, CHOC), 3.33-3.44 (2H, m, CHOC, C$^{3'}$H), 3.40-3.47 (1H, m, CHOCH$_a$H$_b$CH$_2$), 3.64-3.78 (1H, m, CHOCH$_a$H$_b$CH$_2$), 3.98-4.10 (2H, m, CH$_a$H$_b$OAc), 4.26 (2H, d, $J$ = 13.2 Hz, CH$_a$H$_b$OAc), 4.32 (1H, d, $J$ = 7.5 Hz, C$^{1'}$H), 4.41 (1H, d, $J$ = 7.5 Hz, C$^{1''}$H). **$^{13}$C NMR** (100 MHz, DMSO-d$_6$): δ$_C$ 14.4 (CH$_2$CH$_3$), 21.1 (2xCH$_3$C=O), 22.6 (CH$_2$CH$_3$), 25.8 (CH$_2$(CH$_2$)$_2$), 27.0 (2xCH$_2$CH$_2$N), 27.2 (CH$_2$(CH$_2$)$_2$), 27.3 (CH$_2$(CH$_2$)$_2$), 29.3 (CH$_2$(CH$_2$)$_2$), 29.4-29.6 (4xCH$_2$(CH$_2$)$_2$), 29.7 (CHOCH$_2$CH$_2$), 31.7 (CH$_2$(CH$_2$)$_2$), 42.3 (CH$_3$N), 57.6 (2xCH$_2$CH$_2$N), 64.0 (CH$_2$OAc), 64.2 (CH$_2$OAc), 69.0 (CHOCH$_2$CH$_2$), 70.1 (CHOC), 70.2 (CHOC), 73.8 (CHOC), 74.3 (CHOC), 75.2 (C$^{2''}$H), 76.2 (C$^{3'}$H), 76.3 (C$^{3''}$H), 83.0 (C$^{2'}$H), 101.5 (C$^{1'}$H), 104.7 (C$^{1''}$H), 170.6 (CH$_3$C=O), 170.7(CH$_3$C=O).

***N*-methyl-*N*-ocytl-9-[(6',6''-di-*O*-acetyl-2'-*O*-β-D-glucopyranosyl-β-D-glucopyranosyl)oxy]nonan-1-amine (32):** (3.82 g, 70%), beige powder.

**$^1$H NMR** (400 MHz, DMSO-d$_6$): δ$_H$ 0.86 (3H, t, $J$ = 6.7 Hz, CH$_2$CH$_3$), 1.13-1.32 (20H, m, 9xCH$_2$(CH$_2$)$_2$, CH$_2$CH$_3$), 1.33-1.42 (4H, m, 2xCH$_2$CH$_2$N), 1.44-1.56 (2H, m, CHOCH$_2$CH$_2$), 2.00 (3H, s, CH$_3$C=O), 2.01 (3H, s, CH$_3$C=O), 2.08 (3H, s, CH$_3$N), 2.22 (4H, t, $J$ = 7.2 Hz, 2xCH$_2$CH$_2$N), 2.99 (1H, dxd, $J$ = 8.2, 8.2 Hz, C$^{2''}$H), 3.02-3.08 (1H, m, CHOC), 3.08-3.12 (2H, m, 2xCHOC), 3.13-3.17 (2H, m, CHOC, C$^{3''}$H), 3.21 (1H, dxd, $J$ = 9.2, 8.7 Hz, C$^{2'}$H), 3.36-3.40 (1H, m, C$^{3'}$H), 3.40-3.47 (2H, m, CHOCH$_a$H$_b$CH$_2$), 3.50 (1H, dxd, $J$ = 11.9, 4.3 Hz, CH$_a$H$_b$OH), 3.71-3.80 (1H, m, CHOCH$_a$H$_b$CH$_2$) 4.32 (1H, d, $J$ = 7.6 Hz, C$^{1'}$H), 4.40 (1H, d, $J$ = 7.5 Hz, C$^{1''}$H). **$^{13}$C NMR** (100 MHz, DMSO-d$_6$): δ$_C$ 14.4 (CH$_2$CH$_3$), 21.1 (2xCH$_3$C=O), 22.6 (CH$_2$CH$_3$), 25.9 (CH$_2$(CH$_2$)$_2$), 27.1 (2xCH$_2$CH$_2$N), 27.3 (CH$_2$(CH$_2$)$_2$), 27.4 (CH$_2$(CH$_2$)$_2$), 29.2 (CH$_2$(CH$_2$)$_2$), 29.2-29.6 (4xCH$_2$(CH$_2$)$_2$), 29.7 (CHOCH$_2$CH$_2$), 31.7 (CH$_2$(CH$_2$)$_2$), 42.7 (CH$_3$N), 57.6 (2xCH$_2$CH$_2$N), 64.0 (CH$_2$OAc), 64.2 (CH$_2$OAc), 69.1 (CHOCH$_2$CH$_2$), 70.1 (CHOC), 70.2 (CHOC), 73.7 (CHOC), 74.3 (CHOC), 75.2 (C$^{2''}$H), 76.24 (C$^{3'}$H), 76.28 (C$^{3''}$H), 83.0 (C$^{2'}$H), 101.5 (C$^{1'}$H), 104.7 (C$^{1''}$H), 170.70 (CH$_3$C=O), 170.65 (CH$_3$C=O).

***N*-dodecyl-*N*-ocytl-9-[(6',6''-di-*O*-acetyl-2'-*O*-β-D-glucopyranosyl-β-D-glucopyranosyl)oxy]nonan-1-amine (33):** (3.98 g, 73%), beige powder.

**$^1$H NMR** (400 MHz, DMSO-d$_6$): δ$_H$ 0.86 (3H, t, $J$ = 6.9 Hz, CH$_2$CH$_3$), 1.16-1.32 (28H, m, 13xCH$_2$(CH$_2$)$_2$, CH$_2$CH$_3$), 1.32-1.42 (4H, m, 2xCH$_2$CH$_2$N), 1.43-1.55 (2H, m, CHOCH$_2$CH$_2$),

2.01 (3H, s, C$\underline{H}_3$C=O), 2.02 (3H, s, C$\underline{H}_3$C=O), 2.08 (3H, s, C$\underline{H}_3$N), 2.22 (4H, t, $J$ = 6.7 Hz, 2xCH$_2$C$\underline{H}_2$N), 3.01 (1H, dxd, $J$ = 8.5, 8.5 Hz, C$^{2''}\underline{H}$), 3.08-3.19 (3H, m, 2xC$\underline{H}$OC, C$^{3''}\underline{H}$), 3.22 (1H, dxd, $J$ = 7.7, 7.7 Hz, C$^{2'}\underline{H}$), 3.30-3.43 (3H, m, 2xC$\underline{H}$OC, C$^{3'}\underline{H}$), 3.40-3.47 (1H, m, CHOC$\underline{H}_a$H$_b$CH$_2$), 3.76 (1H, dxt, $J$ = 9.0, 6.7 Hz, CHOCH$_a\underline{H}_b$CH$_2$), 3.95-4.10 (2H, m, 2xC$\underline{H}_a$H$_b$OAc), 4.22 (1H, dxd, $J$ = 12.5, 2.0 Hz, CH$_a\underline{H}_b$OAc), 4.25 (1H, dxd, $J$ = 12.2, 2.1 Hz, CH$_a\underline{H}_b$OAc), 4.32 (1H, d, $J$ = 7.6 Hz, C$^{1'}\underline{H}$), 4.41 (1H, d, $J$ = 7.4 Hz, C$^{1''}\underline{H}$), 4.77-5.64 (5H, m, 5xO$\underline{H}$). **$^{13}$C NMR** (100 MHz, DMSO-d$_6$): δ$_C$ 14.4 (CH$_2\underline{C}$H$_3$), 21.1 (2x$\underline{C}$H$_3$C=O), 22.6 ($\underline{C}$H$_2$CH$_3$), 25.9 ($\underline{C}$H$_2$(CH$_2$)$_2$), 27.1 ($\underline{C}$H$_2$CH$_2$N), 27.2 ($\underline{C}$H$_2$CH$_2$N), 27.3 ($\underline{C}$H$_2$(CH$_2$)$_2$), 27.4 ($\underline{C}$H$_2$(CH$_2$)$_2$), 29.1 ($\underline{C}$H$_2$(CH$_2$)$_2$), 29.3-29.6 (8x$\underline{C}$H$_2$(CH$_2$)$_2$), 29.7 (CHOCH$_2\underline{C}$H$_2$), 31.7 ($\underline{C}$H$_2$(CH$_2$)$_2$), 42.3 ($\underline{C}$H$_3$N), 57.5 (CH$_2\underline{C}$H$_2$N), 57.6 (CH$_2\underline{C}$H$_2$N), 64.0 ($\underline{C}$H$_2$OH), 64.2 ($\underline{C}$H$_2$OH), 69.1 (CHO$\underline{C}$H$_2$CH$_2$), 70.1 ($\underline{C}$HOC), 70.2 ($\underline{C}$HOC), 73.7 ($\underline{C}^{2''}$H), 74.3 ($\underline{C}^{3'}$H), 75.2 ($\underline{C}^{3''}$H), 76.2 ($\underline{C}$HOC), 76.3 ($\underline{C}$HOC), 83.0 ($\underline{C}^{2'}$H), 101.5 ($\underline{C}^{1'}$H), 104.8 ($\underline{C}^{1''}$H), 170.6 (CH$_3\underline{C}$=O), 170.7 (CH$_3\underline{C}$=O).

**$N$-methyl-$N$-octadecyl-9-[(6',6''-di-$O$-acetyl-2'-$O$-$\beta$-D-glucopyranosyl-$\beta$-D-glucopyranosyl)oxy]nonan-1-amine (34):** (6.85 g, 75%), off-white powder.

**$^1$H NMR** (400 MHz, DMSO-d$_6$): δ$_H$ 0.83 (3H, t, $J$ = 6.5 Hz, CH$_2$C$\underline{H}_3$), 1.03-1.92 (38H, m, 19xC$\underline{H}_2$(CH$_2$)$_2$), 1.28-1.41 (2H, m, C$\underline{H}_2$CH$_3$), 1.43-1.56 (6H, m, 2xC$\underline{H}_2$CH$_2$N, CHOCH$_2$C$\underline{H}_2$), 2.00 (3H, s, C$\underline{H}_3$C=O), 2.02 (3H, s, C$\underline{H}_3$C=O), 2.05 (3H, s, C$\underline{H}_3$N), 2.22 (4H, t, $J$ = 6.5 Hz, 2xCH$_2$C$\underline{H}_2$N), 2.99 (1H, dxd, $J$ = 8.3, 8.3 Hz, C$^{2''}\underline{H}$), 3.02-3.18 (3H, m, 2xC$\underline{H}$OC, C$^{3''}\underline{H}$), 3.21 (1H, dxd, $J$ = 8.9, 8.9 Hz, C$^{2'}\underline{H}$), 3.27-3.47 (4H, m, 2xC$\underline{H}$OC, C$^{3'}\underline{H}$, CHOC$\underline{H}_a$H$_b$CH$_2$), 3.71-3.81 (1H, m, CHOCH$_a\underline{H}_b$CH$_2$), 3.95-4.08 (2H, m, 2xC$\underline{H}_a$H$_b$OAc), 4.18-4.30 (2H, m, 2xCH$_a\underline{H}_b$OAc), 4.30 (1H, d, $J$ = 7.4 Hz, C$^{1'}\underline{H}$), 4.39 (1H, d, $J$ = 8.0 Hz, C$^{1''}\underline{H}$), 4.88-5.09 (5H, m, 5xO$\underline{H}$). **$^{13}$C NMR** (100 MHz, DMSO-d$_6$): δ$_C$ 14.2 (CH$_2\underline{C}$H$_3$), 21.1 (2x$\underline{C}$H$_3$C=O), 22.6 ($\underline{C}$H$_2$CH$_3$), 26.0 ($\underline{C}$H$_2$(CH$_2$)$_2$), 27.32 (2x$\underline{C}$H$_2$CH$_2$N), 27.40 ($\underline{C}$H$_2$(CH$_2$)$_2$), 27.45 ($\underline{C}$H$_2$(CH$_2$)$_2$), 29.3 ($\underline{C}$H$_2$(CH$_2$)$_2$), 29.3-29.8 (14x$\underline{C}$H$_2$(CH$_2$)$_2$, (CHOCH$_2\underline{C}$H$_2$), 31.8 ($\underline{C}$H$_2$(CH$_2$)$_2$), 42.1 ($\underline{C}$H$_3$N), 57.7 (2xCH$_2\underline{C}$H$_2$N), 64.02 ($\underline{C}$H$_2$OH), 64.20 ($\underline{C}$H$_2$OH), 69.1 (CHO$\underline{C}$H$_2$CH$_2$), 70.1 ($\underline{C}$HOC), 70.2 ($\underline{C}$HOC), 73.8 ($\underline{C}^{2''}$H), 74.3 ($\underline{C}^{3'}$H), 75.2 ($\underline{C}^{3''}$H), 76.22 ($\underline{C}$HOC), 76.28 ($\underline{C}$HOC), 83.2 ($\underline{C}^{2'}$H), 101.6 ($\underline{C}^{1'}$H), 104.9 ($\underline{C}^{1''}$H), 170.47 (CH$_3\underline{C}$=O), 170.51 (CH$_3\underline{C}$=O).

**General procedure for synthesis of non-acetylated sophorolipid quaternary ammonium salts (35), (36), (37), (38), (39), (40).**

In a 10 mL flame dried pressure resistant vial, non-acetylated sophorolipid amine was dissolved in dry acetonitrile. The solution was cooled down to 0 °C and methyl iodide (1.2 eq.) was added. The vial was closed and heated to 80°C. After 24h, the reaction mixture was cooled down to room temperature and the sophorolipid quaternary salt was precipitated. The precipitate was filtered off and dried under reduced pressure. In the absence of precipitation, the reaction mixture was concentrated under reduced pressure and recrystallized from diethyl ether if necessary. The sophorolipid quaternary salts were obtained in high purity without further purification.

**N,N-dimethyl-N-ethyl-9-[(6',6''-di-O-acetyl-2'-O-β-D-glucopyranosyl-β-D-glucopyranosyl)oxy]nonan-1-ammonium iodide (35):** (2.10 g, 31%), brown powder.

**$^1$H NMR** (400 MHz, DMSO-$d_6$): $\delta_H$ 1.06-1.40 (13H, m, 5xCH$_2$(CH$_2$)$_2$, CH$_2$CH$_3$), 1.41-1.58 (2H, m, CHOCH$_2$CH$_2$), 1.55-1.71 (2H, m, CH$_2$CH$_2$N), 2.01 (6H, br s, 2xCH$_3$C=O), 2.96-3.04 (1H, m, C$^{2''}$H), 3.01 (6H, s, 2xCH$_3$N), 3.04-3.08 (1H, m, CHOC), 3.08-3.13 (2H, m, 2xCHOC), 3.12-3.17 (2H, m, CHOC, C$^{3''}$H), 3.17-3.25 (3H, m, C$^{2'}$H, 2H, m, CH$_2$N), 3.28-3.35 (2H, m, CH$_2$N), 3.33-3.39 (1H, m, C$^{3'}$H), 3.41-3.53 (1H, m, CHOCH$_a$H$_b$CH$_2$), 3.57-3.77 (1H, m, CHOCH$_a$H$_b$CH$_2$), 3.97-4.12 (2H, m, 2xCH$_a$H$_b$OAc), 4.25 (2H, d, $J$ = 13.2 Hz, 2xCH$_a$H$_b$OAc), 4.32 (1H, d, $J$ = 7.3 Hz, C$^{1'}$H), 4.41 (1H, d, $J$ = 7.3 Hz, C$^{1''}$H), 4.50 (1H, t, $J$ = 5.7, OH), 4.93-5.15 (2H, m, OH) 5.30 (2H, d, $J$ = 8.6, OH), 5.48 (1H, br s, OH). **$^{13}$C NMR** (100 MHz, DMSO-$d_6$): $\delta_C$ 7.8 (CH$_2$CH$_3$), 20.6 (2xCH$_3$C=O), 21.6 (CH$_2$CH$_2$N), 25.3 (CH$_2$(CH$_2$)$_2$), 25.7 (CH$_2$(CH$_2$)$_2$), 28.4 (CH$_2$(CH$_2$)$_2$), 28.7 (CH$_2$(CH$_2$)$_2$), 29.1 (CH$_2$(CH$_2$)$_2$), 49.4 (2xCH$_3$N), 58.6 (NCH$_2$CH$_3$), 62.4 (CH$_2$CH$_2$N), 68.5 (CHOCH$_2$CH$_2$), 69.5 (CHOC), 69.7 (CHOC), 73.2 (C$^{2''}$H), 73.8 (C$^{3'}$H), 74.6 (C$^{3''}$H), 75.7 (CHOC), 75.8 (CHOC), 82.4 (C$^{2'}$H), 101.1 (C$^{1'}$H), 104.2 (C$^{1''}$H), 170.2 (2xCH$_3$C=O).

**N-butyl-N,N-dimethyl-9-[(6',6''-di-O-acetyl-2'-O-β-D-glucopyranosyl-β-D-glucopyranosyl)oxy]nonan-1-ammonium iodide (36):** (0.40 g, 28%), light brown powder.

**$^1$H NMR** (400 MHz, DMSO-$d_6$): $\delta_H$ 0.94 (3H, t, $J$ = 7.2 Hz, CH$_2$CH$_3$), 1.13-1.40 (12H, m, 5xCH$_2$(CH$_2$)$_2$, CH$_2$CH$_3$), 1.42-1.56 (2H, m, CHOCH$_2$CH$_2$), 1.56-1.75 (4H, m, 2xCH$_2$CH$_2$N), 2.02 (6H, br s, 2xCH$_3$C=O), 2.91-3.08 (1H, m, C$^{2''}$H), 3.01 (6H, s, 2xCH$_3$N), 3.02-3.13 (1H, m, CHOC), 3.13-3.22 (2H, m, 2xCHOC), 3.05-3.20 (2H, m, CHOC, C$^{3''}$H), 3.17-3.28 (5H, m, C$^{2'}$H, 2xCH$_2$N), 3.33-3.44 (1H, m, C$^{3'}$H), 3.40-3.50 (1H, m, CHOCH$_a$H$_b$CH$_2$), 3.69-3.82 (1H, m, CHOCH$_a$H$_b$CH$_2$), 3.97-4.12 (2H, m, 2xCH$_a$H$_b$OAc), 4.25 (2H, d, $J$ = 13.2 Hz,

2xCH$_a$H$_b$OAc), 4.32 (1H, d, $J$ = 7.2 Hz, C$^{1'}$H), 4.41 (1H, dxd, $J$ = 7.1 Hz, C$^{1''}$H), 5.00-5.13 (1H, m, OH), 5.11-5.23 (1H, m, OH), 5.24-5.40 (2H, m, OH), 5.49 (1H, s br, OH). **$^{13}$C NMR** (100 MHz, DMSO-d$_6$): δ$_C$ 14.0 (CH$_2$CH$_3$), 19.7 (CH$_2$CH$_3$), 21.2 (2xCH$_3$C=O), 22.2 (CH$_2$CH$_2$N), 24.2 (CH$_2$CH$_2$N), 25.8 (CH$_2$(CH$_2$)$_2$), 26.2 (CH$_2$(CH$_2$)$_2$), 28.9 (CH$_2$(CH$_2$)$_2$), 29.2 (2xCH$_2$(CH$_2$)$_2$), 29.7 (CHOCH$_2$CH$_2$), 50.4 (2xCH$_3$N), 63.3 (CH$_2$CH$_2$N), 63.5 (CH$_2$CH$_2$N), 63.9 (CH$_2$OAc), 64.2 (CH$_2$OAc), 69.1 (CHOCH$_2$CH$_2$), 70.1 (CHOC), 70.2 (CHOC), 73.7 (C$^{2''}$H), 74.3 (C$^{3'}$H), 75.2 (C$^{3''}$H), 76.2 (CHOC), 76.3 (CHOC), 82.7 (C$^{2'}$H), 101.7 (C$^{1'}$H), 104.5 (C$^{1''}$H) 170.74 (CH$_3$C=O), 170.75 (CH$_3$C=O).

**$N$,$N$-dimethyl-$N$-hexyl-9-[(6',6''-di-$O$-acetyl-2'-$O$-β-$D$-glucopyranosyl-β-$D$-glucopyranosyl)oxy]nonan-1-ammonium iodide (37):** (0.86 g, 43%), light brown powder.

**$^1$H NMR** (400 MHz, DMSO-d$_6$): δ$_H$ 0.88 (3H, t, $J$ = 6.3 Hz, CH$_2$CH$_3$), 1.04-1.41 (16H, m, 7xCH$_2$(CH$_2$)$_2$, CH$_2$CH$_3$), 1.42-1.57 (2H, m, CHOCH$_2$CH$_2$), 1.57-1.74 (4H, m, 2xCH$_2$CH$_2$N), 2.02 (6H, br s, 2xCH$_3$C=O), 2.96-3.06 (1H, m, C$^{2''}$H), 3.00 (6H, s, 2xCH$_3$N), 3.06-3.19 (2H, m, 2xCHOC), 3.12-3.18 (2H, m, CHOC, C$^{3''}$H), 3.18-3.27 (5H, m, C$^{2'}$H, 2xCH$_2$N), 3.32-3.54 (3H, m, C$^{3'}$H, CHOC, CHOCH$_a$H$_b$CH$_2$), 3.58-3.80 (1H, m, CHOCH$_a$H$_b$CH$_2$), 3.97-4.12 (2H, m, 2xCH$_a$H$_b$OAc), 4.25 (2H, d, $J$ = 13.2 Hz, 2xCH$_a$H$_b$OAc), 4.32 (1H, d, $J$ = 6.9 Hz, C$^{1'}$H), 4.42 (1H, dxd, $J$ = 7.5 Hz, C$^{1''}$H), 5.00-5.08 (1H, m, OH), 5.17 (1H, d, $J$ = 2.7, OH), 5.26-5.37 (2H, m, OH) 5.49 (1H, d, $J$ = 3.1, OH). **$^{13}$C NMR** (100 MHz, DMSO-d$_6$): δ$_C$ 13.9 (CH$_2$CH$_3$), 20.7 (2xCH$_3$C=O), 21.63 (CH$_2$CH$_2$N), 21.69 (CH$_2$CH$_2$N) 21.8 (CH$_2$CH$_3$), 25.4 (CH$_2$(CH$_2$)$_2$), 25.8 (2xCH$_2$(CH$_2$)$_2$), 28.5 (CH$_2$(CH$_2$)$_2$), 28.8 (2xCH$_2$(CH$_2$)$_2$), 29.2 (CHOCH$_2$CH$_2$), 30.6 (CH$_2$(CH$_2$)$_2$), 49.9 (2xCH$_3$N), 62.9 (2xCH$_2$CH$_2$N), 63.5 (CH$_2$OAc), 63.7 (CH$_2$OAc), 68.6 (CHOCH$_2$CH$_2$), 69.6 (CHOC), 69.6 (CHOC), 73.2 (C$^{2''}$H), 73.8 (C$^{3'}$H), 74.7 (C$^{3''}$H), 75.7 (CHOC), 75.8 (CHOC), 82.4 (C$^{2'}$H), 101.1 (C$^{1'}$H), 104.1 (C$^{1''}$H), 170.2 (2xCH$_3$C=O).

**$N$,$N$-dimethyl-$N$-octyl-9-[(6',6''-di-$O$-acetyl-2'-$O$-β-$D$-glucopyranosyl-β-$D$-glucopyranosyl)oxy]nonan-1-ammonium iodide (38):** (3.44 g, 60%), light brown powder.

**$^1$H NMR** (400 MHz, DMSO-d$_6$): δ$_H$ 0.86 (3H, t, $J$ = 6.9 Hz, CH$_2$CH$_3$), 1.14-1.40 (20H, m, 9xCH$_2$(CH$_2$)$_2$, CH$_2$CH$_3$), 1.41-1.56 (2H, m, CHOCH$_2$CH$_2$), 1.55-1.72 (4H, m, 2xCH$_2$CH$_2$N), 2.00 (3H, s, CH$_3$C=O), 2.01 (3H, s, CH$_3$C=O), 2.94-3.03 (1H, m, C$^{2''}$H), 3.00 (6H, s, 2xCH$_3$N), 3.06-3.14 (2H, m, 2xCHOC), 3.13-3.19 (2H, m, CHOC, C$^{3''}$H), 3.18-3.28 (5H, m, C$^{2'}$H, 2xCH$_2$N), 3.32-3.44 (3H, m, C$^{3'}$H, CHOC, CHOCH$_a$H$_b$CH$_2$), 3.58-3.79 (1H, m, CHOCH$_a$H$_b$CH$_2$), 3.96-4.11 (2H, m, 2xCH$_a$H$_b$OAc), 4.25 (2H, m, 2xCH$_a$H$_b$OAc), 4.32 (1H, d,

$J$ = 7.4 Hz, C$^{1'}$<u>H</u>), 4.41 (1H, d, $J$ = 7.6 Hz, C$^{1''}$<u>H</u>), 5.00-5.08 (1H, m, O<u>H</u>), 5.17 (1H, d, $J$ = 5.6, O<u>H</u>), 5.25-5.40 (2H, m, O<u>H</u>) 5.50 (1H, d, $J$ = 3.6, O<u>H</u>). **$^{13}$C NMR** (100 MHz, DMSO-d$_6$): δ$_C$ 14.4 (CH$_2$<u>C</u>H$_3$), 21.2 (2x<u>C</u>H$_3$C=O), 22.1 (2x<u>C</u>H$_2$CH$_2$N), 22.6 (<u>C</u>H$_2$CH$_3$), 25.9 (<u>C</u>H$_2$(CH$_2$)$_2$), 26.3 (2x<u>C</u>H$_2$(CH$_2$)$_2$), 28.9 (4x<u>C</u>H$_2$(CH$_2$)$_2$), 29.2 (<u>C</u>H$_2$(CH$_2$)$_2$), 29.7 (CHOCH$_2$<u>C</u>H$_2$), 31.6 (<u>C</u>H$_2$(CH$_2$)$_2$), 50.5 (2x<u>C</u>H$_3$N), 63.4 (2xCH$_2$<u>C</u>H$_2$N), 64.0 (<u>C</u>H$_2$OAc), 64.2 (<u>C</u>H$_2$OAc), 69.1 (CHO<u>C</u>H$_2$CH$_2$), 70.0 (<u>C</u>HOC), 70.2 (<u>C</u>HOC), 73.7 (<u>C</u>$^{2''}$H), 74.3 (<u>C</u>$^{3'}$H), 75.2 (<u>C</u>$^{3''}$H), 76.2 (<u>C</u>HOC), 76.3 (<u>C</u>HOC), 82.9 (<u>C</u>$^{2'}$H), 101.5 (<u>C</u>$^{1'}$H), 104.6 (<u>C</u>$^{1''}$H), 170.7 (2xCH$_3$<u>C</u>=O).

**_N,N_-dimethyl-_N_-dodecyl-9-[(6',6''-di-_O_-acetyl-2'-_O_-β-D-glucopyranosyl-β-D-glucopyranosyl)oxy]nonan-1-ammonium iodide (39):** (4.02 g, 74%), beige powder.

**$^1$H NMR** (400 MHz, DMSO-d$_6$): δ$_H$ 0.86 (3H, t, $J$ = 6.6 Hz, CH$_2$C<u>H</u>$_3$), 1.10-1.40 (28H, m, 13xC<u>H</u>$_2$(CH$_2$)$_2$, C<u>H</u>$_2$CH$_3$), 1.42-1.56 (2H, m, CHOCH$_2$C<u>H</u>$_2$), 1.57-1.70 (4H, m, 2xC<u>H</u>$_2$CH$_2$N), 2.02 (6H, br s, 2xC<u>H</u>$_3$C=O), 2.96-3.03 (1H, m, C$^{2''}$<u>H</u>), 2.99 (6H, s, 2xC<u>H</u>$_3$N), 3.03-3.08 (1H, m, C<u>H</u>OC), 3.08-3.12 (2H, m, 2xC<u>H</u>OC), 3.12-3.18 (2H, m, C<u>H</u>OC, C$^{3''}$<u>H</u>), 3.18-3.27 (5H, m, C$^{2'}$<u>H</u>, 2xC<u>H</u>$_2$N), 3.32-3.54 (2H, m, C$^{3'}$<u>H</u>, CHOC<u>H</u>$_a$H$_b$CH$_2$), 3.63-3.78 (1H, m, CHOCH$_a$<u>H</u>$_b$CH$_2$), 3.96-4.12 (2H, m, 2xCH$_a$<u>H</u>$_b$OAc), 4.26 (2H, d, $J$ = 12.2 Hz, 2xC<u>H</u>$_a$H$_b$OAc), 4.32 (1H, d, $J$ = 7.7 Hz, C$^{1'}$<u>H</u>), 4.41 (1H, dxd, $J$ = 7.3 Hz, C$^{1''}$<u>H</u>). **$^{13}$C NMR** (100 MHz, DMSO-d$_6$): δ$_C$ 14.4 (CH$_2$<u>C</u>H$_3$), 21.2 (2x<u>C</u>H$_3$C=O), 21.1 (2x<u>C</u>H$_2$CH$_2$N), 22.6 (<u>C</u>H$_2$CH$_3$), 25.9 (<u>C</u>H$_2$(CH$_2$)$_2$), 26.2 (2x<u>C</u>H$_2$(CH$_2$)$_2$), 28.9 (<u>C</u>H$_2$(CH$_2$)$_2$), 29.1-29.6 (8x<u>C</u>H$_2$(CH$_2$)$_2$), 29.7 (CHOCH$_2$<u>C</u>H$_2$), 31.8 (<u>C</u>H$_2$(CH$_2$)$_2$), 50.5 (2x<u>C</u>H$_3$N), 63.4 (2xCH$_2$<u>C</u>H$_2$N), 64.0 (<u>C</u>H$_2$OAc), 64.2 (<u>C</u>H$_2$OAc), 69.1 (CHO<u>C</u>H$_2$CH$_2$), 70.0 (<u>C</u>HOC), 70.2 (<u>C</u>HOC), 73.7 (<u>C</u>$^{2''}$H), 74.3 (<u>C</u>$^{3'}$H), 75.2 (<u>C</u>$^{3''}$H), 76.2 (<u>C</u>HOC), 76.6 (<u>C</u>HOC), 82.9 (<u>C</u>$^{2'}$H), 101.5 (<u>C</u>$^{1'}$H), 104.7 (<u>C</u>$^{1''}$H), 170.7 (2xCH$_3$<u>C</u>=O).

**_N,N_-dimethyl-_N_-octadecyl-9-[(6',6''-di-_O_-acetyl-2'-_O_-β-D-glucopyranosyl-β-D-glucopyranosyl)oxy]nonan-1-ammonium iodide (40):** (1.90 g, 35%), off-white powder.

**$^1$H NMR** (400 MHz, DMSO-d$_6$): δ$_H$ 0.86 (3H, t, $J$ = 6.9 Hz, CH$_2$C<u>H</u>$_3$), 1.00-1.42 (40H, m, 18xC<u>H</u>$_2$(CH$_2$)$_2$, C<u>H</u>$_2$CH$_3$), 1.43-1.55 (2H, m, CHOCH$_2$C<u>H</u>$_2$), 1.55-1.76 (4H, m, 2xC<u>H</u>$_2$CH$_2$N), 2.02 (6H, br s, 2xC<u>H</u>$_3$C=O), 2.96-3.08 (1H, m, C$^{2''}$<u>H</u>), 3.01 (6H, s, 2xC<u>H</u>$_3$N), 3.02-3.10 (1H, m, C<u>H</u>OC), 3.08-3.15 (2H, m, 2xC<u>H</u>OC), 3.15-3.24 (2H, m, C<u>H</u>OC, C$^{3''}$<u>H</u>), 3.17-3.33 (4H, m, 2xC<u>H</u>$_2$N), 3.28-3.37 (1H, m, C$^{2'}$<u>H</u>), 3.38-3.55 (2H, m, C$^{3'}$<u>H</u>, CHOC<u>H</u>$_a$H$_b$CH$_2$), 3.66-3.83 (1H, m, CHOCH$_a$<u>H</u>$_b$CH$_2$), 3.98-4.10 (2H, m, 2xCH$_a$<u>H</u>$_b$OAc), 4.26 (2H, d, $J$ = 12.2 Hz, 2xC<u>H</u>$_a$H$_b$OAc), 4.32 (1H, d, $J$ = 7.6 Hz, C$^{1'}$<u>H</u>), 4.40 (1H, dxd, $J$ = 7.7, 2.7 Hz, C$^{1''}$<u>H</u>). **$^{13}$C**

**NMR** (100 MHz, DMSO-d$_6$): δ$_C$ 14.4 (CH$_2$$\underline{C}$H$_3$), 21.1 (2x$\underline{C}$H$_3$C=O), 21.2 (2x$\underline{C}$H$_2$CH$_2$N), 22.5 ($\underline{C}$H$_2$CH$_3$), 25.9 ($\underline{C}$H$_2$(CH$_2$)$_2$), 26.2 (2x$\underline{C}$H$_2$(CH$_2$)$_2$), 28.9 ($\underline{C}$H$_2$(CH$_2$)$_2$), 29.1-29.6 (14x$\underline{C}$H$_2$(CH$_2$)$_2$), 29.7 (CHOCH$_2$$\underline{C}$H$_2$), 31.8 ($\underline{C}$H$_2$(CH$_2$)$_2$), 50.4 (2x$\underline{C}$H$_3$N), 63.4 (2xCH$_2$$\underline{C}$H$_2$N), 64.0 ($\underline{C}$H$_2$OAc), 64.2 ($\underline{C}$H$_2$OAc), 69.1 (CHO$\underline{C}$H$_2$CH$_2$), 70.0 ($\underline{C}$HOC), 70.2 ($\underline{C}$HOC), 73.6 ($\underline{C}$$^{2''}$H), 74.3 ($\underline{C}$$^{3'}$H), 75.2 ($\underline{C}$$^{3''}$H), 76.2 ($\underline{C}$HOC), 76.6 ($\underline{C}$HOC), 82.9 ($\underline{C}$$^{2'}$H), 101.6 ($\underline{C}$$^{1'}$H), 104.8 ($\underline{C}$$^{1''}$H), 170.7 (2xCH$_3$$\underline{C}$=O).

Table S 1 – List of all sample studied in this work ("Sample N° in manuscript column" refers to numbers of samples listed in Figure 2, Figure 5, Figure 8, Figure 10 in main manuscript), names, Mw, dry matter, appearance, solubility and cmc. Abbreviations: *Acetyl*: Acetylated; *Bola*: Bolaform; *Non-Sym*: Non-symmetrical; *Sym*: symmetrical; *Lact*: Lactonic. *: CMC values are given both in wt% and in mM units. Values inside brackets () are given in mM.

| Sample N° | Surfactant Name | Mw [g/mol] | Dry matter [%] | Appearance | Solubility [wt%] | CMC* [wt%] (mM) |
|---|---|---|---|---|---|---|
| Bolaform Sophoro-*Y* ; *Y*= -Side (SS) ; *Y*= -Lipid (SL) ||||||||
| (1) | (Acetyl) Bola SL C18:1 [ω, ω-1] | ~1000 | 100 | Powder | pH 4: 0.1 ≤ S < 0.2<br>pH 6: > 2<br>pH 8: >2 | pH 4: 0.067 (0.717)<br>pH 6: 0.068 (0.728)<br>pH 8: 0.025 (0.268) |
| (2) | Acetyl Bola SS C18:0 [ω] | 1102 | 100 | Powder | INS < 0.1 | / |
| (3) | Acetyl Bola SS C18:1, [ω] | 1100 | 95-99% | Powder | S >2 | pH 4: 0.006 (0.055)<br>pH 6: 0.012 (0.110)<br>pH 8: 0.011 (0.101) |
| (4) | Acetyl Bola SS C18:1, [ω, ω-1] | 1100 | 100 | Powder | - | / |
| (5) | Non-Acetyl Bola SS C18:1 [ω] | 932 | 100 | Powder | S < 0.1 | / |
| Sophorolipids (SL) C18:*X* (*X*= 0, 1) ||||||||
| (6) | Acetyl Acidic SL C18:1 [ω] | 706 | 38.3% | Viscous solution | pH 4 : 0 ≤ S < 0.1<br>pH 6 : 1 ≤ S < 2<br>pH 8 : S > 2 | / |
| (7) | Acetyl Acidic SL C18:1 [ω-1] | 706 | 35.6% | Viscous solution | S > 1.4 | pH 4: 0.042 (0.591)<br>pH 6: 0.034 (0.483)<br>pH 8: 0.107 (1.517) |
| (8) | Non-Acetyl Acidic SL C18:1 [ω-1] (minor amount [ω]) | 622 | 100 | Powder | pH 4 : S < 0.1<br>pH 6 : 0.1 < S <0.2<br>pH 8 : S > 2 | pH 4 : / / / /<br>pH 6 : 0.039 (0.626)<br>pH 8 : 0.045 (0.722) |
| (9) | Acetyl Lact SL C18:1 [ω-1] (minor amount [ω]) | 688 | 100 | Powder | INS < 0.1 | / |
| (10) | Acetyl Lact SL C18:0 [ω-1] | 690 | 100 | Powder | INS | / |
| (11) | Acetyl Acidic SL C18:0 [ω-1] | 708 | 100 | Powder | pH 4 : < 0.1<br>pH 6 : < 0.1<br>pH 8 : 0.1 < S < 0.2 | / |
| | Glucolipids (GL) [ω-1]; C18:X (X= 0, 1) | | | | | |
| (12) | Acetyl GL C18:1 | 502 | 72.1% | Viscous solution | pH 4 : < 0.08<br>pH 6 : < 0.08<br>pH 8 : 0.8 < S <1.6 | pH 4 : / / / /<br>pH 6 : / / / /<br>pH 8 : 0.013 (0.248) |
| (13) | Acetyl Acidic GL C18:0 | 504 | 100 | Powder | INS < 0.1 | / |
| | Sophoro-Y C9:0 [ω] ; Y= -Side (SS) ; Y= -Lipid (SL) | | | | | |
| (14) | Acetyl Acidic SL C9:0 | 582 | 100 | Powder | 1 < S < 2 | pH 4 : 0.162 (2.778)<br>pH 6 : 0.203 (3.481)<br>pH 8 : 0.312 (5.350) |
| (15) | Acetyl Aldehyde SS C9:0 | 566 | 100 | Powder | S > 2 | pH 4 : 0.078 (1.375)<br>pH 6 : 0.023 (10.41)<br>pH 8 : 0.100 (1.763 ) |
| (16) | Acetyl Alcohol SS C9:0 | 568 | 100 | Powder | S > 2 | pH 4 : 0.068 (1.195)<br>pH 6 : 0.098 (1.722)<br>pH 8 : 0.166 (2.917) |
| (17) | Non-Acetyl Acidic SL C9:0 | 498 | 100 | Powder | S > 2 | / |
| (18) | Non-Acetyl Alcohol SS C9:0 | 484 | 100 | Powder | 0.1 < S < 0.2 | pH 4 : / / / /<br>pH 6 : / / / /<br>pH 8 : 0.672 (13.86) |
| (19) | Non-Acetyl Amine SS C9:0-C2 | 525 | 100 | Powder | pH 4 : 0.1 < S < 0.2<br>pH 6 : S < 0.1<br>pH 8 : 1 < S < 2 | / |
| (20) | Non-Acetyl Amine SS C9:0-C4 | 553 | 100 | Powder | pH 4 : 0.2 < S < 0.5<br>pH 6 : S < 0.1<br>pH 8 : 0.2 < S < 0.5 | pH 4 : / / / /<br>pH 6 : / / / /<br>pH 8 : 3.476 (62.73) |
| (21) | Non-Acetyl Amine SS C9:0-C6 | 581 | 100 | Powder | pH 4 : 0.5 < S < 1<br>pH 6 : S > 2<br>pH 8 : S > 2 | pH 4 : / / / /<br>pH 6 : 0.515 (8.846)<br>pH 8 : 0.256 (4.397) |
| (22) | Non-Acetyl Amine SS C9:0-C8 | 609 | 100 | Powder | pH 4 : S > 2<br>pH 6 : S < 0.1<br>pH 8 : S > 2 | pH 4 : 0.798 (13.077)<br>pH 6 : / / / /<br>pH 8 : 0.12 (1.967) |
| (23) | Non-Acetyl Amine SS C9:0-C12 | 665 | 100 | Powder | S > 2 | pH 4 : 0.024 (0.360)<br>pH 6 : 0.049 (0.735<br>pH 8 : 0.024 (0.360) |
| (24) | Non-Acetyl Amine SS C9:0-C18 | 749 | 100 | Powder | pH 4 : 0.2 < S < 0.5 | pH 4 : 0.010 (0.133) |

| | | | | | pH 6 : 0.2 < S < 0.5<br>pH 8 : S < 0.1 | pH 6 : / / / /<br>pH 8 : / / / / |
|---|---|---|---|---|---|---|
| (25) | Non-Acetyl Ammonium SS C9:0-C2 | 666 | 100 | Powder | pH 4: S < 0.1<br>pH 6: S < 0.1<br>pH 8: 1 < S < 2 | pH 4 : / / / /<br>pH 6 : / / / /<br>pH 8 : 0.826 (12.378) |
| (26) | Non-Acetyl Ammonium SS C9:0-C4 | 694 | 100 | Powder | pH 4: S < 0.1<br>pH 6: S < 0.1<br>pH 8: 0.2 < S < 0.5 | pH 4 : / / / /<br>pH 6 : / / / /<br>pH 8 : 1.051 (15.114) |
| (27) | Non-Acetyl Ammonium SS C9:0-C6 | 722 | 100 | Powder | pH 4: S < 0.1<br>pH 6: S < 0.1<br>pH 8: 0.2 < S < 0.5 | pH 4 : / / / /<br>pH 6 : / / / /<br>pH 8 : 1.055 (14.583) |
| (28) | Non-Acetyl Ammonium SS C9:0-C8 | 750 | 100 | Powder | pH 4: S < 0.1<br>pH 6: S < 0.1<br>pH 8: 0.2 < S < 0.5 | pH 4 : / / / /<br>pH 6 : / / / /<br>pH 8 : 0.803 (10.685) |
| (29) | Acetyl Amine SS C9:0-C2 | 609 | 100 | Powder | S > 2 | / |
| (30) | Acetyl Amine SS C9:0-C4 | 637 | 100 | Powder | S > 2 | / |
| (31) | Acetyl Amine SS C9:0-C6 | 665 | 100 | Powder | S > 2 | / |
| (32) | Acetyl Amine SS C9:0-C8 | 693 | 100 | Powder | S > 2 | / |
| (33) | Acetyl Amine SS C9:0-C12 | 749 | 100 | Powder | S > 2 | / |
| (34) | Acetyl Amine SS C9:0-C18 | 834 | 100 | Powder | pH 4: 0 < S < 0.1<br>pH 6: 0.1 < S < 0.2<br>pH 8: 0 < S < 0.1 | / |
| | | | | | | / |
| (35) | Acetyl Ammoniun SS C9:0-C2 | 750 | 100 | Powder | S > 2 | / |
| (36) | Acetyl Ammonium SS C9:0-C4 | 778 | 100 | Powder | S > 2 | / |
| (37) | Acetyl Ammoniun SS C9:0-C6 | 806 | 100 | Powder | 1 < S < 2 | / |
| (38) | Acetyl Ammoniun SS C9:0-C8 | 834 | 100 | Powder | S > 2 | / |
| (39) | Acetyl Ammoniun SS C9:0-C12 | 890 | 100 | Powder | S > 2 | / |
| (40) | Acetyl Ammoniun SS C9:0-C18 | 975 | 100 | Powder | 0 < S < 0.1 | / |

**Surface tension data**

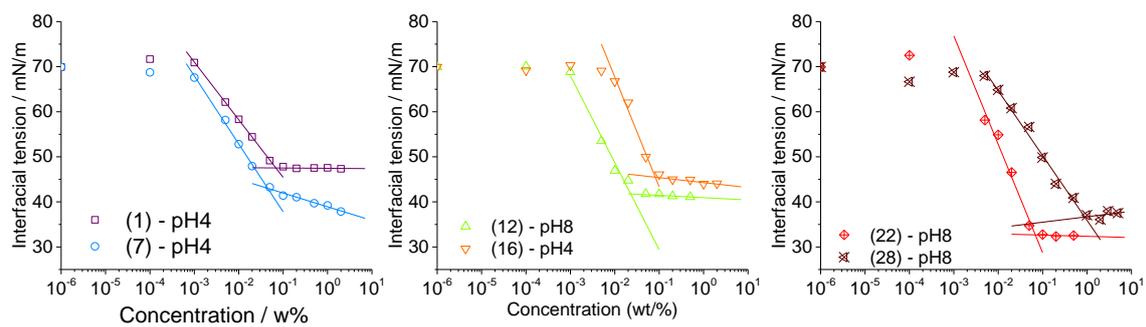

**Figure S 1 – Surface tension versus concentration plots and corresponding linear fits used to determine the CMC values given in Table S 1. CMC is determined at the intersection of the linear fits. Full table containing surface tension data is shown in Table S 2.**

**Table S 2 – Surface tension data recorded as a function of concentration**

| Sample | (1) | (1) | (1) | (3) | (3) | (7) | (7) | (7) | (8) | (8) | (12) | (14) | (14) | (14) | (15) | (15) | (15) | (16) | (16) | (16) | (18) |
|---|---|---|---|---|---|---|---|---|---|---|---|---|---|---|---|---|---|---|---|---|---|
| C /wt% | pH4 | pH 6 | pH 8 | pH 6 | pH 8 | pH4 | pH 6 | pH 8 | pH 6 | pH 8 | pH 8 | pH4 | pH 6 | pH 8 | pH4 | pH 6 | pH 8 | pH4 | pH 6 | pH 8 | pH 8 |
| $10^{-6}$ | 69.93 | 69.93 | 69.93 | 69.93 | 69.93 | 69.93 | 69.93 | 69.93 |  | 72.14 | 69.93 | 69.93 | 69.93 | 69.93 | 69.93 | 69.93 | 69.93 | 69.93 | 69.93 | 69.93 | 69.93 |
| $10^{-5}$ |  |  |  |  |  |  |  |  |  |  |  |  |  |  |  |  |  |  |  |  |  |
| $10^{-4}$ | 71.67 | 69.52 | 71.01 |  | 53.23 | 73.17 | 68.73 | 71.7 | 70.71 |  | 70.03 | 72.01 | 71.03 | 70.75 | 71.58 | 70.56 | 68.96 | 69.17 | 66.29 | 72.45 | 72.32 |
| $10^{-3}$ | 70.91 | 64.79 | 67.76 | 70.51 | 57.80 | 70.2 | 67.57 | 70.79 | 69.92 | 71.56 | 68.83 | 72.62 | 69.97 | 71.17 | 71.52 | 70.68 | 69.75 | 70.3 | 67 | 72.15 | 72.41 |
| $5.10^{-3}$ | 62.15 | 63.44 | 61.78 | 63.06 | 49.38 | 59.56 | 58.16 | 60.95 | 66 | 62.82 | 53.52 | 74.16 | 70.06 | 71.07 | 70.74 | 68.44 | 69.15 | 69.08 | 68.45 | 69.24 | 72.31 |
| $10^{-2}$ | 58.32 | 59.92 | 55.23 | 47.12 | 44.46 | 53.8 | 52.81 | 57.64 | 58.7 | 57.2 | 46.93 | 72.27 | 69.62 | 70.27 | 65.53 | 63.43 | 65.33 | 66.74 | 68.75 | 68.09 | 65.06 |
| $2 10^{-2}$ | 54.42 | 56.1 | 50.09 | 44.42 | 43.43 | 46.97 | 47.93 | 54.2 | 51.65 | 51.6 | 44.76 | 69.8 | 67.44 | 69.25 | 58.57 | 47.33 | 55.74 | 62.03 | 65.12 | 65.21 | 65.66 |
| $5.10^{-2}$ | 49.16 | 51.83 | 48.02 | 43.8 | 43.73 | 40.99 | 43.25 | 49.2 | 44.64 | 48.2 | 41.78 | 59.13 | 59.52 | 60.74 | 48.88 | 44.52 | 51.86 | 49.97 | 53.7 | 55.63 | 62.88 |
| $10^{-1}$ | 47.8 | 49.88 | 47.43 | 43.54 | 43.26 | 39.75 | 41.40 | 45.74 | 42.87 | 46.26 | 41.75 | 50.12 | 49.97 | 53.38 | 45.55 | 44.19 | 45.23 | 46.08 | 47.7 | 50.43 | 59.92 |
| $2.10^{-1}$ | 47.44 | 49.44 | 47.68 | 42.99 | 43.71 | 39.23 | 41.04 | 44.04 | 42.69 | 46.01 | 41.29 | 46.05 | 46 | 47.55 | 44.12 | 43.18 | 44.26 | 44.97 | 45.58 | 46.96 | 57.01 |
| $5.10^{-1}$ | 47.52 | 48.86 | 47.19 | 43.08 | 43.18 | 38.86 | 39.69 | 42.77 | 42.41 | 45.76 | 41.09 | 43.4 | 42.94 | 43.81 | 42.85 | 42.53 | 43.34 | 44.9 | 45.02 | 45.18 | 53.94 |
| 1 | 47.55 | 48.77 | 47.15 | 42.34 | 42.61 | 38.77 | 39.21 | 42.38 | 41.69 | 45.66 |  | 43.17 | 41.64 | 42.74 | 42.37 | 41.6 | 42.83 | 43.97 | 45.05 | 45.22 | 52.3 |
| 2 | 47.34 | 48.00 | 46.58 | 42.39 | 42.35 | 38.42 | 37.86 | 41.86 | 41.87 | 45.01 |  | 42.24 | 42.16 | 42.55 | 41.73 | 41.14 | 41.79 | 44.05 | 44.86 | 45.09 | 51.7 |
| 3 |  |  |  |  |  |  |  |  |  |  |  |  |  |  |  |  |  |  |  |  |  |
| 5 |  |  |  |  |  |  |  |  |  |  |  |  |  |  |  |  |  |  |  |  |  |

**Table S 2 – continued**

| Sample C /wt% | (20) pH 4 | (20) pH 8 | (21) pH4 | (21) pH 6 | (21) pH 8 | (22) pH 4 | (22) pH 8 | (23) pH4 | (23) pH 6 | (23) pH 8 | (24) pH 4 | (25) pH 8 | (26) pH 8 | (27) pH 8 | (28) pH 8 |
|---|---|---|---|---|---|---|---|---|---|---|---|---|---|---|---|
| $10^{-7}$ | | | | | | | | 68.11 | 69.93 | 69.93 | | | | | |
| $10^{-6}$ | 69.93 | 69.93 | 69.93 | 69.93 | 69.93 | 69.93 | 69.93 | 67.29 | 73.43 | 70.00 | 69.93 | 69.93 | 69.93 | 69.93 | 69.93 |
| $10^{-5}$ | | | | | | | | 67.84 | 73.19 | 70.89 | | | | | |
| $10^{-4}$ | 73.03 | 71.41 | 61.76 | 70.57 | 69.91 | 71.96 | 71.77 | 54.54 | 73.17 | 70.53 | 66.46 | 70.91 | 67.82 | 67.53 | 66.66 |
| $10^{-3}$ | 73.3 | 72.73 | 62.27 | 70.18 | 69.02 | 73.03 | 70.56 | 49.47 | 63.1 | 65.33 | 69.6 | 70.04 | 66.97 | 68.02 | 68.77 |
| $5.10^{-3}$ | 72.21 | 69.81 | 60.08 | 68.5 | 67.15 | 66.92 | 66.27 | 46.97 | 51.1 | 46.32 | 62.24 | 69.9 | 69.04 | 65.99 | 67.98 |
| $10^{-2}$ | 69.16 | 68.29 | 61.76 | 64.86 | 64.77 | 64.52 | 62.38 | 42.76 | 46.58 | 42.43 | 41.32 | 66.8 | 66.42 | 64.75 | 64.87 |
| $210^{-2}$ | 66.48 | 65.73 | 60.08 | 61.65 | 60.08 | 59.28 | 57.57 | 37.87 | 39.98 | 35.79 | 40.77 | 64.44 | 62.81 | 61.94 | 60.81 |
| $5.10^{-2}$ | 62.09 | 60.97 | 58.05 | 56.4 | 54.61 | 54.34 | 48.12 | 36.19 | 34.88 | 33.59 | 40.18 | 60.03 | 59.53 | 57.33 | 56.64 |
| $10^{-1}$ | 58.23 | 57.92 | 55.73 | 52.83 | 46.99 | 49.18 | 35.79 | | 34.89 | 32.81 | 39.67 | 56.7 | 56.88 | 54.21 | 49.85 |
| $2.10^{-1}$ | 55.14 | 55.22 | 53.43 | 48.65 | 37.27 | 43.07 | 32.31 | 35.58 | | 32.84 | 39.34 | 53.52 | 54.54 | 51.91 | 43.99 |
| $5.10^{-1}$ | 51.62 | 50.31 | 49.98 | 42.25 | 33.95 | 35.89 | 32.99 | 35.19 | 34.78 | | 38.8 | 50.31 | 51 | 47.2 | 40.89 |
| 1 | 47.12 | 46.89 | 45.56 | 43.48 | 35.07 | 32.11 | 32.29 | 34.36 | 35.83 | 32.28 | 38.46 | 48.48 | 49.01 | 44.74 | 37.05 |
| 2 | 44.78 | 40.54 | 38.84 | 41.04 | 33.66 | 32.53 | 32.26 | 34.5 | | | 38.27 | 47.44 | 47.74 | 43.15 | 36.09 |
| 3 | 43.13 | 35.52 | | | | | | | | | | 48.61 | 48.42 | 43.74 | 37.94 |
| 5 | 41.76 | 35.37 | | | | | | | | | | 48.5 | 47.38 | 42.5 | 37.52 |

**SAXS data recorded at concentrations between 0.12 wt% and 16.7 wt% for samples (1)-(40)**

The *.dat* files corresponding to all SAXS profiles shown in Figure S 1 are freely available at https://doi.org/10.5281/zenodo.7677356 (all files are grouped in a single zip file) for further use and analysis.

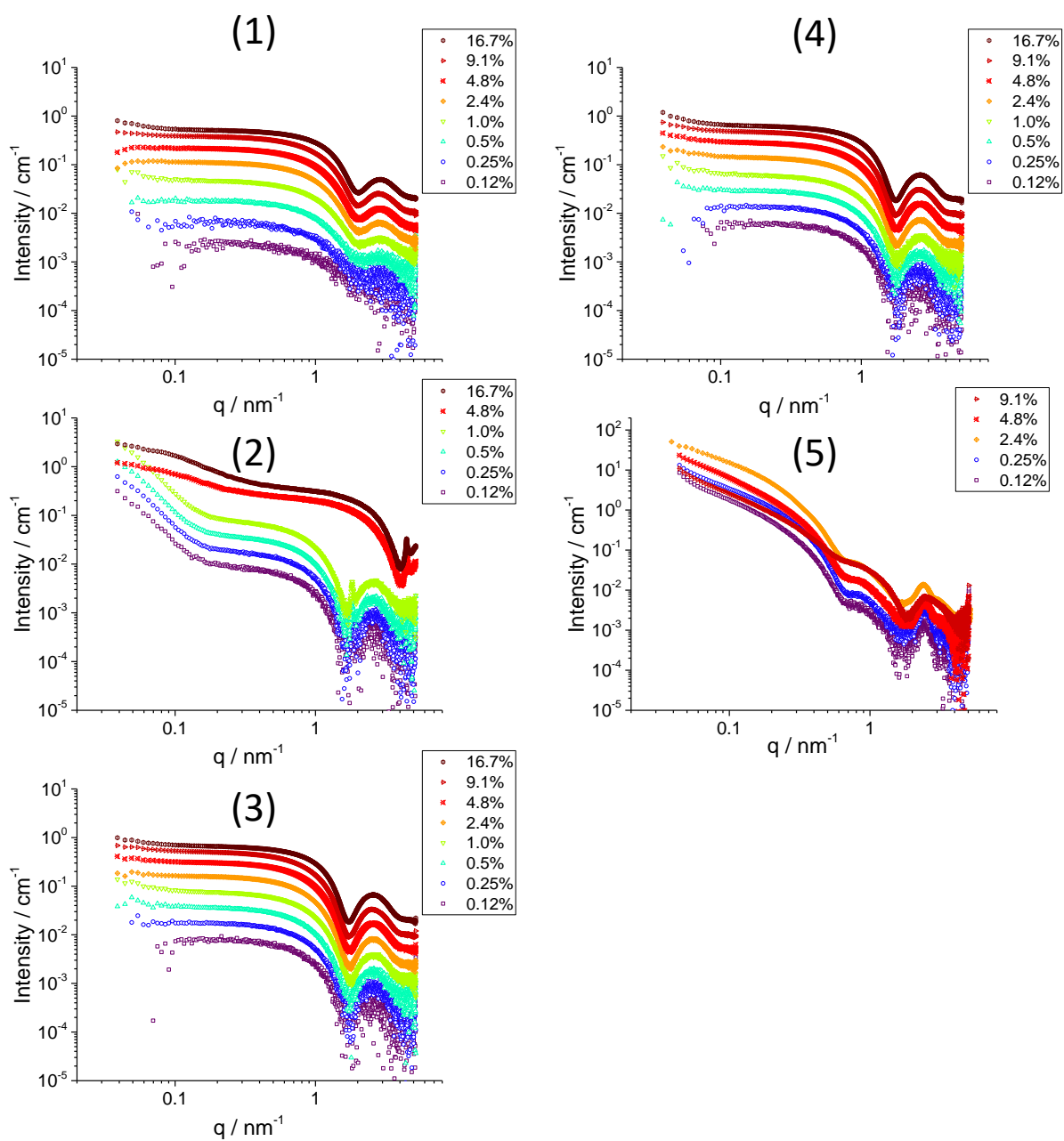

**Figure S 2** – SAXS profiles recorded on microbial glycolipids in water and concentration range between 0.1 wt% and 16 wt%. Chemical formulas (1) to (40) are given both in the main text and on Page S2 in the Supporting Information.

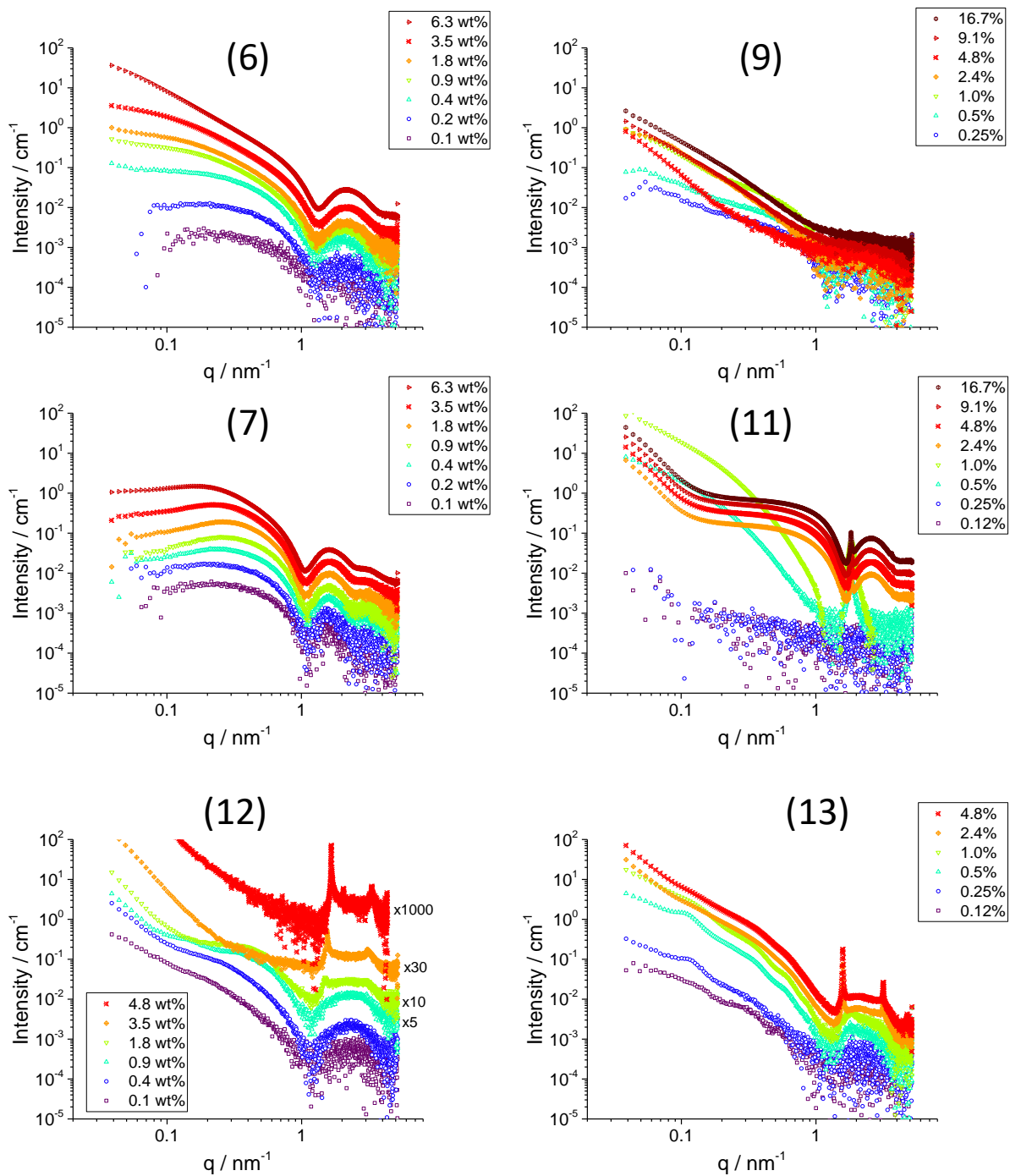

**Figure S 1 (continued)** - SAXS profiles recorded on microbial glycolipids in water and concentration range between 0.1 wt% and 16 wt%. Chemical formulas (1) to (40) are given both in the main text and on Page S2 in the Supporting Information.

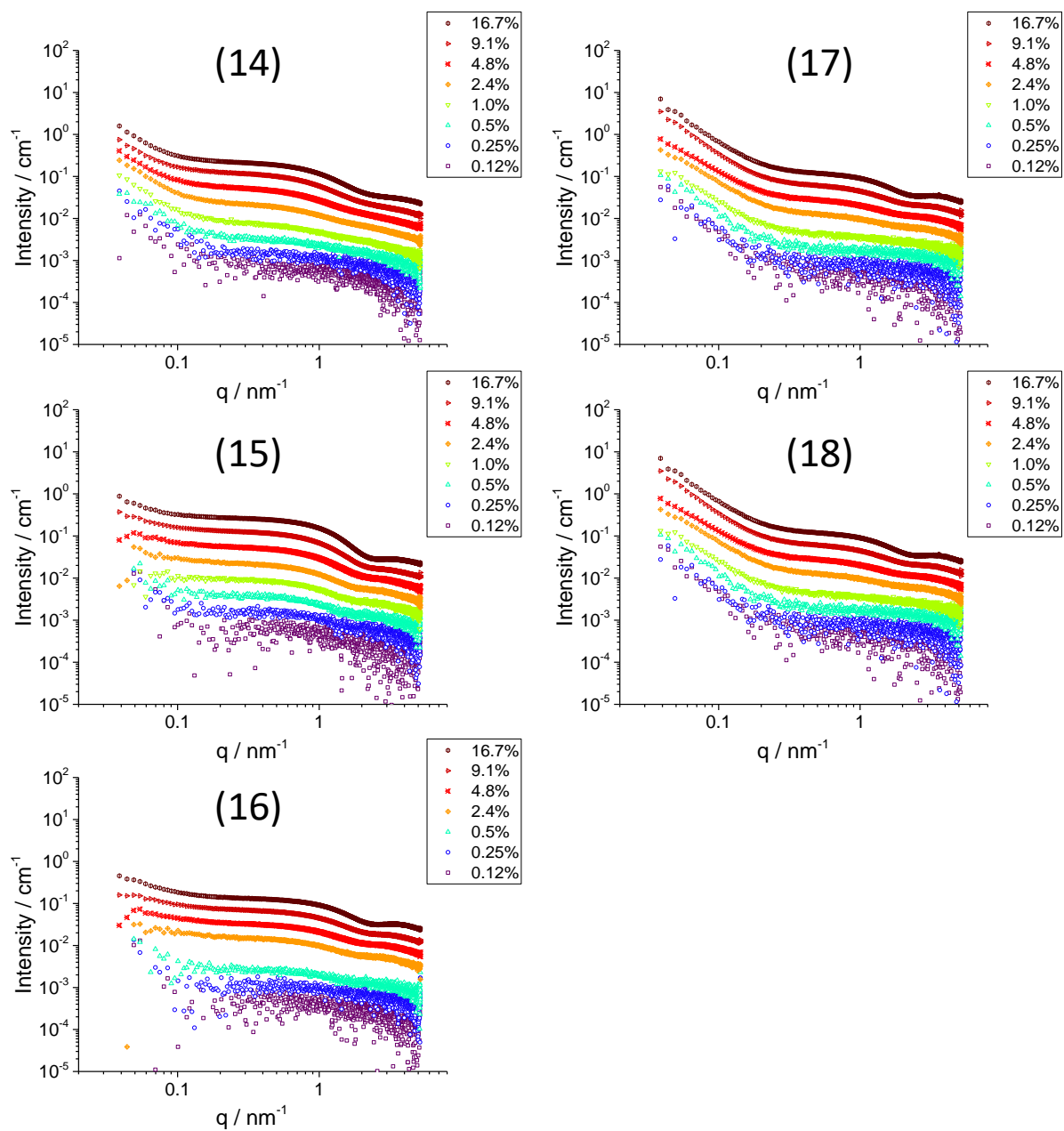

**Figure S 1 (continued) - SAXS profiles recorded on microbial glycolipids in water and concentration range between 0.1 wt% and 16 wt%. Chemical formulas (1) to (40) are given both in the main text and on Page S2 in the Supporting Information.**

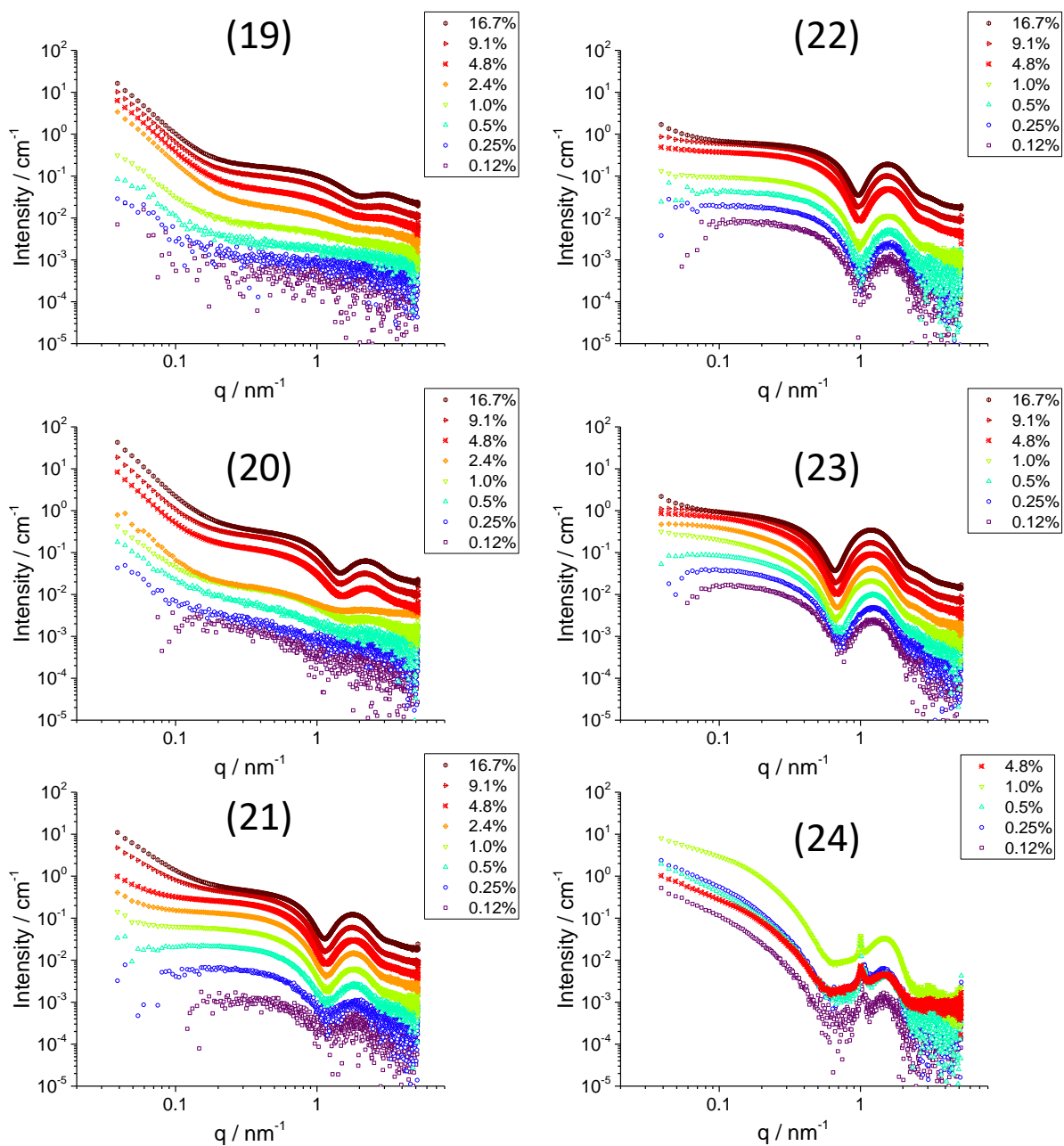

**Figure S 1 (continued)** - SAXS profiles recorded on microbial glycolipids in water and concentration range between 0.1 wt% and 16 wt%. Chemical formulas (1) to (40) are given both in the main text and on Page S2 in the Supporting Information.

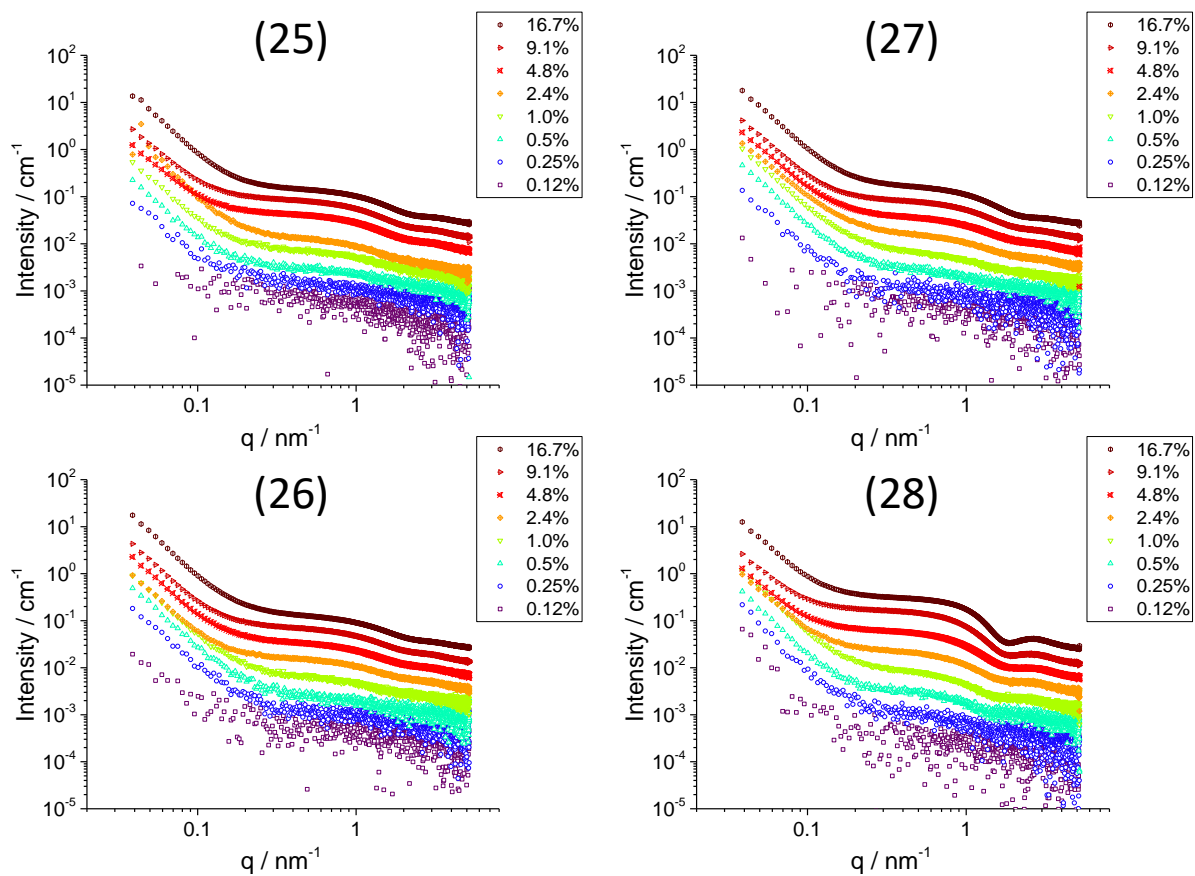

**Figure S 1 (continued) - SAXS profiles recorded on microbial glycolipids in water and concentration range between 0.1 wt% and 16 wt%. Chemical formulas (1) to (40) are given both in the main text and on Page S2 in the Supporting Information.**

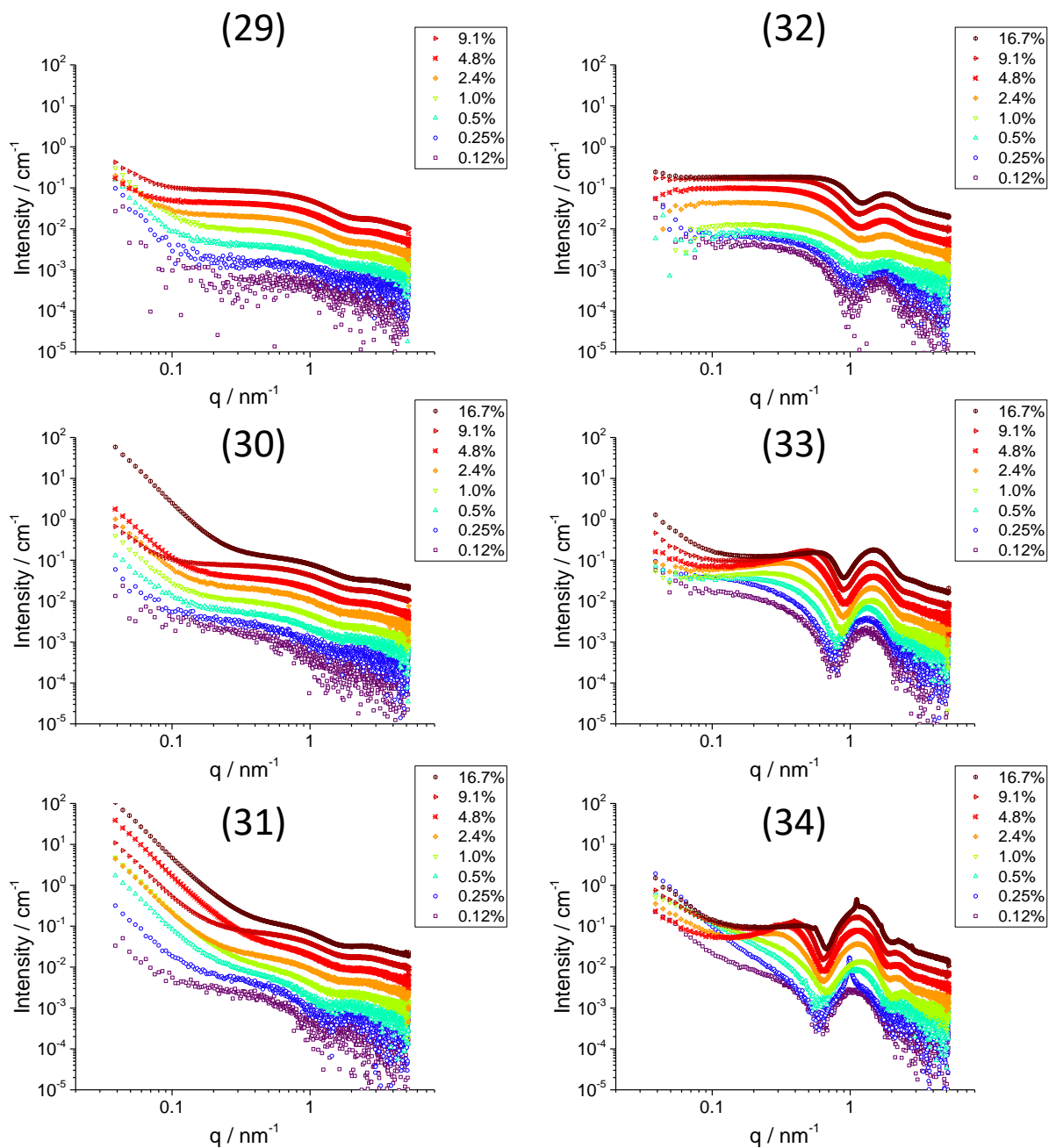

**Figure S 1 (continued) -** SAXS profiles recorded on microbial glycolipids in water and concentration range between 0.1 wt% and 16 wt%. Chemical formulas (1) to (40) are given both in the main text and on Page S2 in the Supporting Information.

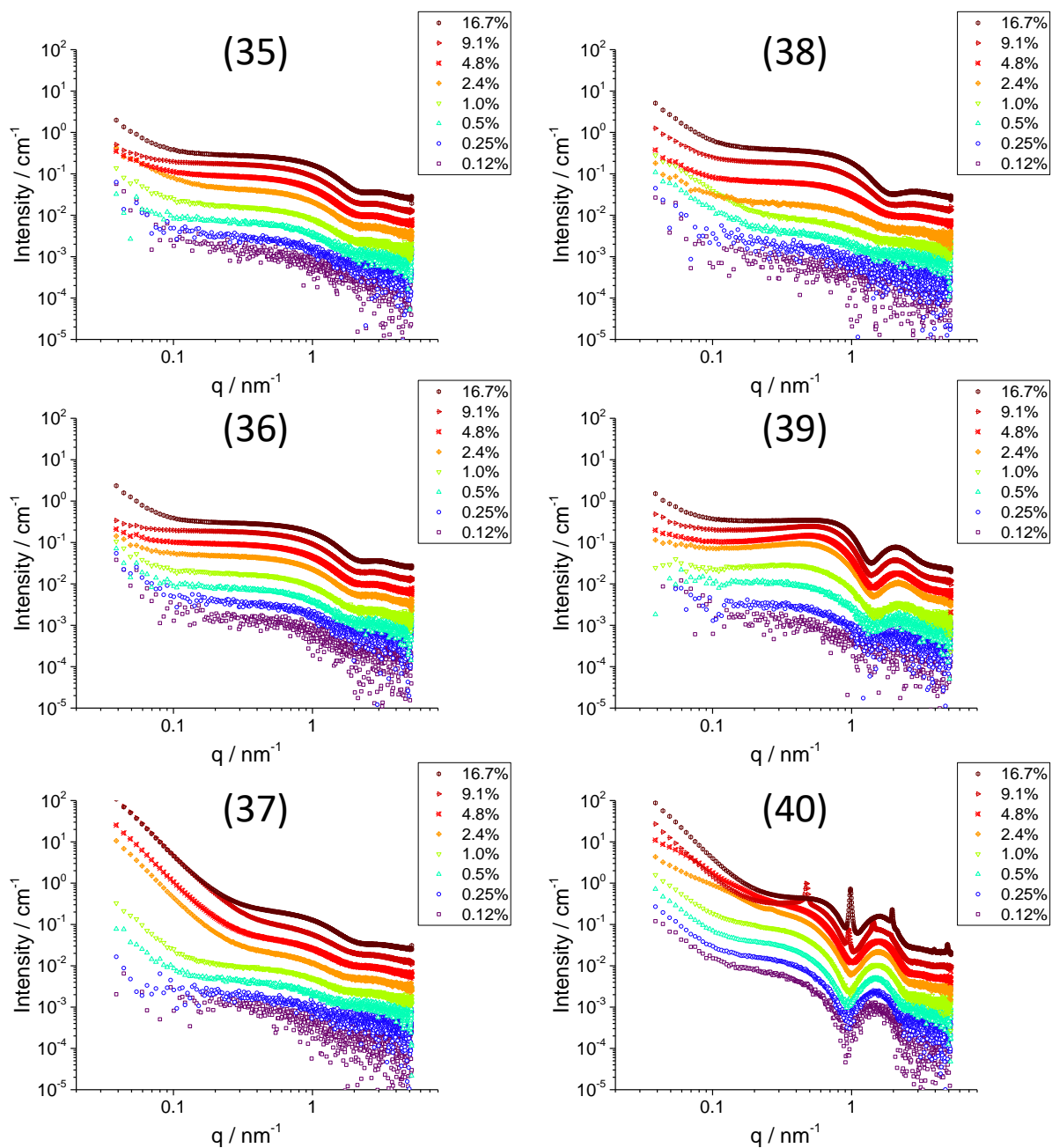

**Figure S 1 (continued)** - SAXS profiles recorded on microbial glycolipids in water and concentration range between 0.1 wt% and 16 wt%. Chemical formulas (1) to (40) are given both in the main text and on Page S2 in the Supporting Information.

# Microscopy

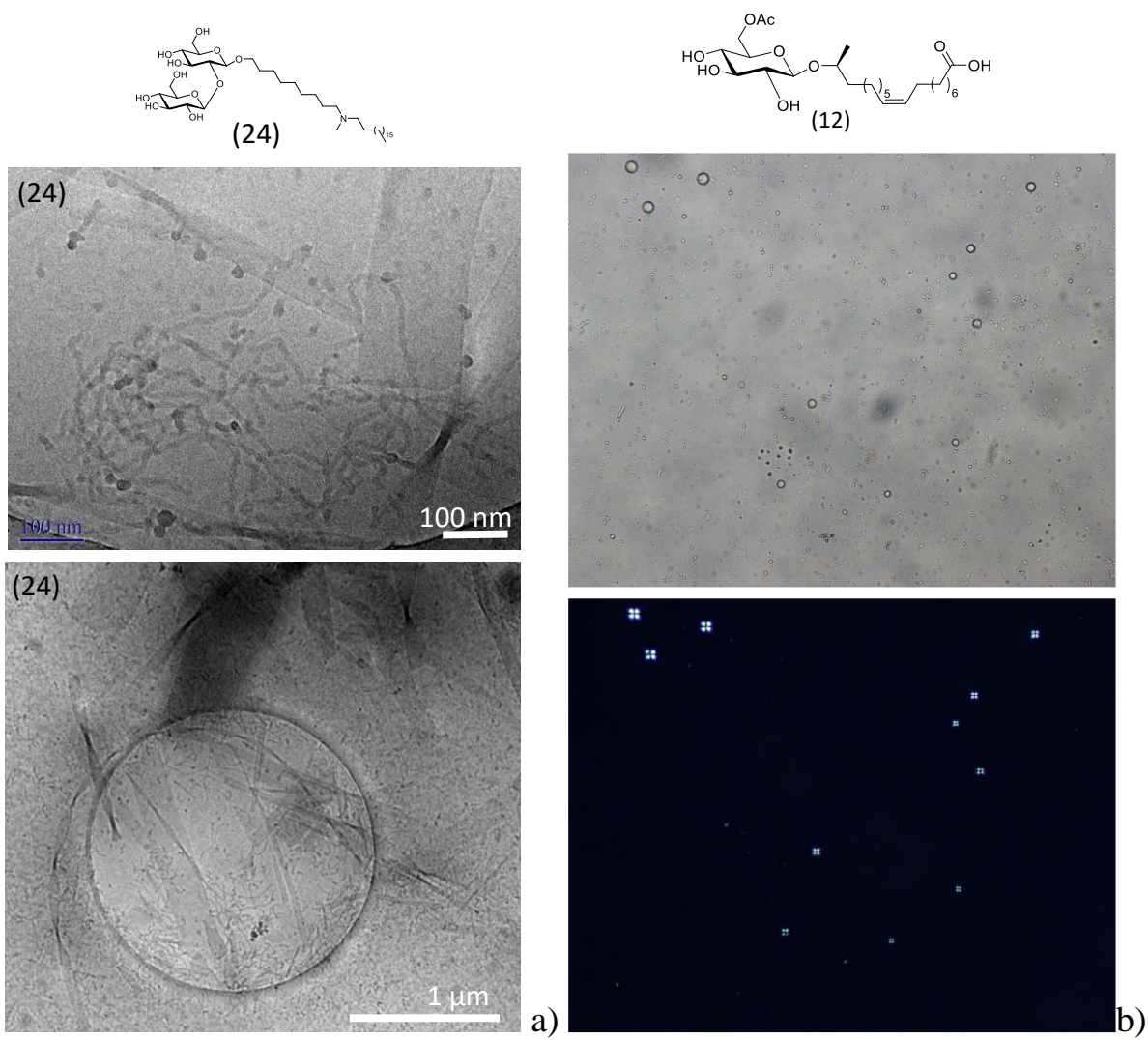

**Figure S 3** – a) Cryo-TEM (C(24)= 0.25 wt%) and b) polarized light microscopy (C(12)= 1 wt%) of aqueous solutions of sophorolipids.